\documentclass[fleqn,usenatbib]{mnras}
\usepackage{newtxtext,newtxmath}
\usepackage[T1]{fontenc}
\usepackage{booktabs}

\DeclareRobustCommand{\VAN}[3]{#2}
\let\VANthebibliography\thebibliography
\def\thebibliography{\DeclareRobustCommand{\VAN}[3]{##3}\VANthebibliography}

\usepackage{graphicx} 
\usepackage{amsmath}
\usepackage[normalem]{ulem}
\usepackage{comment}
\usepackage{multirow}

\title[The dwarf stellar mass function]{The dwarf stellar mass function in different environments and the lack of a generic missing dwarfs problem in $\Lambda$CDM}

\author[I. Lazar et al.]{I. Lazar\thanks{E-mail: i.lazar@herts.ac.uk},$^{1}$ S. Kaviraj,$^{1}$ G. Martin,$^{2}$ A. Watkins,$^{1}$ D. Kakkad,$^{1}$ B. Bichang'a,$^{1}$ K. Kraljic,$^{3}$ S. K. Yi,$^{4}$ \newauthor Y. Dubois,$^{5}$ J. E. G. Devriendt,$^{6}$ S. Peirani$^{7,8}$ and C. Pichon$^{5,9}$\\
$^{1}$Centre for Astrophysics Research, Department of Physics, Astronomy and Mathematics, University of Hertfordshire, College Lane, Hatfield AL10 9AB, UK\\
$^{2}$School of Physics and Astronomy, University of Nottingham, University Park, Nottingham NG7 2RD, UK\\
$^{3}$Observatoire Astronomique de Strasbourg, Universit\'e de Strasbourg, CNRS, UMR 7550, F-67000 Strasbourg, France\\
$^{4}$Department of Astronomy and Yonsei University Observatory, Yonsei University, 50 Yonsei-ro, Seodaemun-gu, Seoul 03722, Republic of Korea\\
$^{5}$Institut d'Astrophysique de Paris, Sorbonne Universit\'es, UMPC Univ Paris 06 et CNRS, UMP 7095, 98 bis bd Arago, 75014 Paris, France\\
$^{6}$Department of Physics, University of Oxford, Keble Road, Oxford OX1 3RH UK\\
$^{7}$ILANCE, CNRS – University of Tokyo International Research Laboratory, Kashiwa, Chiba 277-8582, Japan\\
$^{8}$Kavli IPMU (WPI), UTIAS, The University of Tokyo, Kashiwa, Chiba 277-8583, Japan\\
$^{9}$Kyung Hee University, Dept. of Astronomy \& Space Science, Yongin-shi, Gyeonggi-do 17104, Republic of Korea
}

\pubyear{\the\year{}}

\begin{document}
\label{firstpage}
\pagerange{\pageref{firstpage}--\pageref{lastpage}}
\maketitle

\begin{abstract}
We combine deep photometric data in the COSMOS and XMM-LSS fields with high-resolution cosmological hydrodynamical simulations to explore two key questions: (1) how does the galaxy stellar mass function, particularly in the dwarf ($M_\star$ $<$ 10$^{9.5}$ M$_\odot$) regime, vary with environment, defined as distance from the large-scale structure (LSS) traced by nodes and filaments in the cosmic web? (2) is there a generic `missing dwarfs' problem in $\Lambda$CDM predictions when all environments -- and not just satellites around Milky Way like galaxies -- are considered? The depth of the observational data used here enables us to construct complete, unbiased samples of galaxies, down to $M_{\star}$ $\sim$ 10$^7$ M$_{\odot}$ and out to $z\sim0.4$. Strong environmental differences are found for the galaxy stellar mass function when considering distance from LSS. As we move closer to LSS, the dwarf mass function becomes progressively flatter and the knee of the mass function shifts to larger stellar masses, both of which result in a higher ratio of massive to dwarf galaxies. While the stellar mass functions from the three simulations (\textsc{NewHorizon}, TNG50 and \textsc{FIREbox}) considered here do not completely agree across the dwarf regime, there is no evidence of a generic missing dwarfs problem in the context of $\Lambda$CDM, akin to the results of recent work that demonstrates that there is no missing satellites problem around Galactic analogues. 
\end{abstract}

\begin{keywords}
{\color{black}galaxies: formation -- galaxies: evolution -- galaxies: luminosity function, mass function -- galaxies: dwarf} 
\end{keywords}


\section{Introduction} 

{\color{black}In the hierarchical $\Lambda$CDM paradigm dwarf ($M_\star$ $<$ 10$^{9.5}$ M$_\odot$) galaxies represent an important population of astrophysical objects. Their numerical dominance enables this population to bring about macroscopic evolutionary changes, such as making a large contribution to the re-ionization of the Universe \citep[e.g.][]{Salvadori2015,Atek2015}. The shallow gravitational potential wells of these systems makes them sensitive laboratories for studying both internal processes, such as baryonic feedback \citep[][]{Dong2003,Martin2025}, and external, environmentally-driven processes like tidal perturbations \citep[e.g.][]{Koch2012,Martel2012,Fattahi2018,Jackson2021a,Watkins2023} and ram pressure stripping \citep[e.g.][]{Mori2000,Mayer2006,Steyrleithner2020,Boselli2022}. Their high ratio of dark to baryonic mass also makes dwarf galaxies good testbeds for the study of dark matter \citep[e.g.][]{Battaglia2022,Jackson2021b}. 

Despite their significance, the dwarf population remains much more poorly understood than massive galaxies. In the nearby Universe, dwarfs have been studied in detail largely in our local neighbourhood (and in relatively high-density environments), such as in the Local Group \citep[e.g.][]{Tolstoy2009}, around nearby massive galaxies \citep[e.g.][]{Poulain2021,Trujillo2021,Geha2024} or in nearby galaxy clusters \citep[e.g.][]{Aguerri2005,Boselli2008,ChoqueChallapa2021}. Such studies have either employed deep targeted observations or have observed regions which are close enough that relatively complete samples of dwarfs can be assembled using the data available from existing surveys.   

It is worth noting, however, that typical dwarfs are not bright enough to be detectable outside the local Universe in past wide area surveys like the Sloan Digital Sky Survey (SDSS), which offer large footprints (that facilitate statistical studies at low redshift) but are relatively shallow. Dwarfs that do appear in such datasets and reside outside the local neighbourhood have high star formation rates (SFRs), which are needed to boost their luminosities above the detection limits of the shallow images. Thus, while large samples of dwarfs do exist in surveys like the SDSS, they represent a biased, blue star-forming subset of the dwarf population \citep[e.g.][]{Kaviraj2025}. This bias may lead to erroneous conclusions, such as underestimated red/quenched fractions, fractions of active galactic nuclei (AGN) that are too low (because the strong star formation may swamp the signatures of an AGN) and a morphological mix that is likely skewed towards late-type galaxies (because late-types typically host stronger star formation). 

Assembling unbiased, statistical samples of dwarfs in low density environments has traditionally been challenging because it requires surveys that are both deep and wide. While this will become routinely possible in the new era of surveys like LSST \citep[e.g.][]{Ivezic2019} and Euclid \citep{Laureijs2011,Cuillandre2025}, some deep multi-wavelength datasets do now exist which enable such studies, albeit over relatively small areas of the sky. The ability to access large, complete samples of dwarf galaxies outside the local Universe offers the opportunity for unbiased explorations of many properties (e.g. star formation, quenching, morphology and the presence of AGN) that we were, until recently, limited to exploring largely in massive galaxies only. For example, in a series of recent studies, we have used the deep photometric data available in the 2 deg$^2$ COSMOS field \citep{Scoville2007} to explore the properties of the dwarf population outside the very local Universe \citep{Lazar2024a,Lazar2024b,Bichanga2024,Kaviraj2025}. Notwithstanding the relatively modest area of a field like COSMOS, the shape of the galaxy mass function\footnote{The mass function measures the number density of galaxies as a function of stellar mass.} means that complete, statistical samples of dwarfs can be constructed in these parts of the sky out to moderate redshift ($z\sim0.4)$. 

These papers have demonstrated interesting differences between the properties of dwarfs and massive galaxies, which often diverge from the results of past dwarf studies underpinned by shallow surveys, due to the biases described above. For example, while the well-known `early-type' and `late-type' morphological classes that dominate the Hubble sequence \citep{Hubble1936} in massive galaxies also exist in dwarfs, a progressively larger fraction of galaxies appears to be `featureless' towards lower stellar masses in the dwarf regime \citep{Lazar2024a}. This featureless class is akin to galaxies which are described as `dwarf spheroidals' near the Milky Way and in nearby clusters \citep[e.g.][]{shapley38,reaves56,vandenbergh59,Kormendy1985}. In contrast to their massive counterparts, dwarf early-types exhibit a lower incidence of interactions, are less concentrated, show positive colour gradients and share similar rest-frame colours as dwarf late-types \citep{Lazar2024b}. This suggests that their formation histories are likely to be shaped by secular processes (rather than by interactions) and that they grow `outside in' rather than `inside out', in contrast with their massive counterparts. 

Interactions in dwarfs appear to enhance star formation across the entire body of the galaxy \citep{Lazar2024b}, rather than in the galaxy centres, as is often the case in massive galaxies. Finally, contrary to what has been suggested by dwarf studies that are underpinned by shallow surveys like the SDSS, the AGN fractions in dwarfs appear similar (or perhaps even higher) than those in massive galaxies (e.g. \citealp[][ see also \citealp{Mezcua2024}]{Bichanga2024}). This suggests that the role of AGN in dwarfs -- which remains poorly understood -- may be important in the low-mass regime \citep[e.g.][]{Silk2017,Dashyan2018,Koudmani2021,Koudmani2022,Davis2022}. Together with the recent literature, these studies demonstrate that significant advances in our understanding of galaxy evolution can be made by studying complete, unbiased samples of galaxies in the dwarf regime. 

In this paper, we use deep, multi-wavelength data to probe the galaxy stellar mass function in the nearby Universe down to $M_\star$ $\sim$ 10$^{7}$ M$_\odot$ and out to $z\sim0.4$, which represents the parameter space within which mass-complete samples of galaxies can be constructed, given the depth of the available data (see \citet{Kaviraj2025} and the discussion in Section \ref{sec:completeness}). We first study how the galaxy stellar mass function varies with environment, defined as the distance from large-scale structure (LSS) traced by the nodes and filaments in the cosmic web. We then compare our observed mass functions to those predicted by high-resolution cosmological hydrodynamical simulations (\textsc{NewHorizon}, TNG50 and \textsc{FIREbox}), to explore whether there is a `missing dwarfs' problem in $\Lambda$CDM predictions when galaxies across all environments are considered. This is a generalised version of the `missing satellites' problem, which is a longstanding potential tension between the predicted and observed number densities of low mass galaxies in the standard paradigm \citep[e.g.][]{Klypin1999,Moore1999,Boylan-Kolchin2011,Bullock2017}. The issue of missing dwarfs has so far been probed in satellite populations around the Milky Way or other local massive galaxies. However, if such a problem does exist, it will not be restricted just to the Milky Way neighbourhood but rather manifest itself as a generic missing \textit{dwarfs} problem across all environments. The deep data now available enables just such an extended exploration of missing dwarfs outside the local neighbourhood, albeit restricted to larger stellar masses than has been probed in the context of the missing satellites problem.  

The plan for this paper is as follows. In Section \ref{sec:data}, we describe the datasets used in this study, the algorithm used for calculating environmental parameters and the cosmological hydrodynamical simulations that we compare to the mass functions derived from the observational data. In Section \ref{sec:completeness}, we calculate the lower stellar mass and upper redshift limits out to which we can construct complete galaxy samples, given the depth of the observational data. In Section \ref{sec:mass_function_uncertainties}, we describe how the uncertainties on mass functions are calculated. In Section \ref{sec:mass_functions}, we study how the stellar mass function, particularly in the dwarf regime, varies with distance from LSS traced by the nodes and filaments that define the cosmic web. In Section \ref{sec:missing_dwarfs} we probe whether there is a generic missing dwarfs problem as described above. We summarise our findings in Section \ref{sec:summary}. {\color{black} Throughout this work we assume a Hubble constant ($H_0$) of 70.5 km s$^{-1}$ Mpc$^{-1}$.} 


\section{Data}
\label{sec:data}

{\color{black}In this study we use deep multi-wavelength observational survey data in the COSMOS and XMM-LSS fields, in conjunction with three high-resolution cosmological hydrodynamical simulations. In the sections below we describe each dataset and the extraction of environmental parameters and briefly outline the characteristics of the simulations.} 


\subsection{The COSMOS field}

\label{sec:cos20}

{\color{black}We use the Classic version of the COSMOS2020 catalogue \citep{Weaver2022}, which provides physical parameters, such as photometric redshifts, stellar masses and SFRs for around 1.7 million sources in the $\sim$2 deg$^2$ COSMOS \citep{Scoville2007} field (centered at 10h, +02$^{\circ}$). The physical parameters are calculated using deep photometry in around 40 broad and medium band filters, from the UV through to the mid-infrared, from the following instruments: GALEX \citep{Zamojski2007}, MegaCam/CFHT \citep{Sawicki2019}, ACS/HST \citep{Leauthaud2007}, Hyper Suprime-Cam \citep{Aihara2019}, Subaru/Suprime-Cam \citep{Taniguchi2007,Taniguchi2015}, VIRCAM/VISTA \citep{McCracken2012} and IRAC/Spitzer \citep{Ashby2013,Steinhardt2014,Ashby2015,Ashby2018}. 

The detection image used for object identification in COSMOS2020 includes optical ($i,z$) images from the Ultradeep layer of the HSC Subaru Strategic Program (HSC-SSP), which has a point-source depth of $\sim$28 mag \citep{Aihara2019}. As a comparison, this is $\sim$10 mag fainter than the magnitude limit of the SDSS spectroscopic main galaxy sample \citep[e.g.][]{Alam2015}. The optical and infrared aperture photometry is extracted using the \textsc{SExtractor} \citep{Bertin1996} and \textsc{IRACLEAN} \citep{Hsieh2012} codes respectively. Parameter estimation is then performed using the \textsc{LePhare} SED-fitting algorithm \citep{Arnouts2002,Ilbert2006} run on all broad and medium band filters from UV through to IRAC Channel 2 (which has a central wavelength of 4.5 $\mu$m). The deep data, combined with the large number of photometric filters, results in photometric redshift accuracies better than $\sim$1 and $\sim$4 per cent for bright ($i<22.5$ mag) and faint ($25<i<27$ mag) galaxies respectively. In our analysis below, we use the redshifts and stellar masses directly from the COSMOS2020 catalogue.}  


\subsection{The XMM-LSS field}

{\color{black}We use deep $u$-band to near-infrared photometry, produced using SExtractor, from the CLAUDS XMM-LSS catalogue \citep{Desprez2023,Picouet2023}, which is based on data within a 4.1 deg$^2$ footprint in the XMM-LSS field (centered at 02h, -04$^{\circ}$) from the following instruments: MegaCam/CFHT \citep{Sawicki2019}, Hyper Suprime-Cam \citep{Aihara2019} and VIRCAM/VISTA \citep{McCracken2012}. 

The detection image used for object identification in this dataset includes optical ($i,z$) images from the Deep layer of the HSC-SSP, which has a point-source depth of $\sim$27 mag (1 mag shallower than the HSC-SSP images in COSMOS). Photometric redshifts and stellar masses are calculated in an identical fashion to COSMOS2020 via \textsc{LePhare}, using the broadband $ugrizyYJHK$ photometry available in this catalogue. The accuracy of the physical parameters are comparable to the ones obtained for the COSMOS2020 catalogue. The different locations and sizes of the COSMOS and XMM-LSS fields aids in reducing the uncertainties related to cosmic variance when considering the galaxy stellar mass function.} 


\subsection{DiSPerSE: measuring environmental parameters}

\label{sec:disperse}

{\color{black}In the sections below, we study how the stellar mass function varies with environment, defined as the distance from LSS. To perform this analysis, we measure the projected distances of our dwarf galaxies from the nodes and filaments that define the cosmic web using the DisPerSE algorithm \citep{Sousbie2011}. We follow \citet{Lazar2023} and \citet{Bichanga2024}, who have performed an identical density analysis using the COSMOS2020 catalogue. DisPerSE starts by  producing a density map using Delaunay tessellations, which is calculated using the spatial positions of galaxies \citep{Schaap2000}. Stationary (critical) points in the density map (i.e. the maxima, minima and saddles) correspond to local nodes, voids and the centres of filaments respectively. Segments are used to connect the local maxima with saddle points, forming a set of ridges that defines the network of filaments that describes the cosmic web. A `persistence' parameter ($N$) is used to set a threshold value for defining pairs of critical points that are used to produce the final density map. Only critical pairs with Poisson probabilities above \textit{N}$\sigma$ from the mean are retained. Following \citet{Lazar2023} and \citet{Bichanga2024} we use a persistence value of 2 in our study, which removes ridges close to the noise level, where structures could be spurious.

We use massive ($M_{\star}$ > 10$^{10}$ M$_{\odot}$) galaxies to construct the density maps because they exhibit the smallest redshift errors ($\delta z \sim0.008$ at $z\sim0.2$) and dominate their local gravitational potential wells. The accuracy of the COSMOS2020 and CLAUDS XMM-LSS redshifts enables us to employ well-defined and relatively narrow redshift slices within which we build our 2D density maps. When constructing each map, individual galaxies are weighted by the area under their redshift probability density function that is contained within the slice in question. This ensures that the uncertainties in the photometric redshifts are propagated through the construction of the maps. Once the maps have been constructed and the stationary points located, we calculate the projected distances from the nearest nodes and filaments for each galaxy in our catalogues. Finally, as noted in \citet{Lazar2023}, the low number counts of massive galaxies at $z<0.2$ makes it difficult to create density maps at these redshifts. Our analysis of how the mass function varies as a function of location in the cosmic web therefore focuses on the redshift range $0.2<z<0.4$. Figure \ref{fig:density_map} shows an example density map from our analysis.

We complete this section by considering the types of environments that are likely to be present in our COSMOS and XMM-LSS footprints in the redshift range $0.2<z<0.4$, by considering the \textit{M}$_{200}$ values (which are proxies for the virial masses) of groups identified in the literature \citep{Finoguenov2007,George2011,Gozaliasl2014,Gozaliasl2019}. The virial masses of groups in COSMOS lie in the range 10$^{13}$ M$_{\odot}$ < \textit{M}$_{\rm{200}}$ < 10$^{14.5}$ M$_{\odot}$, with a median value of 10$^{13.4}$ M$_{\odot}$. The virial masses of groups in XMM-LSS lie in the range 10$^{13.3}$ M$_{\odot}$ < \textit{M}$_{\rm{200}}$ < 10$^{14.1}$ M$_{\odot}$, with a similar median value of 10$^{13.6}$ M$_{\odot}$. For comparison, a small cluster like Fornax has a virial mass of $\sim$10$^{13.9}$ M$_{\odot}$ \citep{Drinkwater2001}, while large clusters like Virgo and Coma have virial masses of $\sim$10$^{15}$ M$_{\odot}$ \citep[e.g.][]{Fouque2001,Gavazzi2009}. The galaxy population considered in this study therefore resides mainly in relatively low-density environments (i.e. groups and the field) and not in rich clusters. It is worth noting that the simulations used in Section \ref{sec:missing_dwarfs} bracket the types of environments in the observations and are also not focused on clusters, making the comparisons between data and theory consistent.}  

\begin{figure}
\center
\includegraphics[width=\columnwidth]{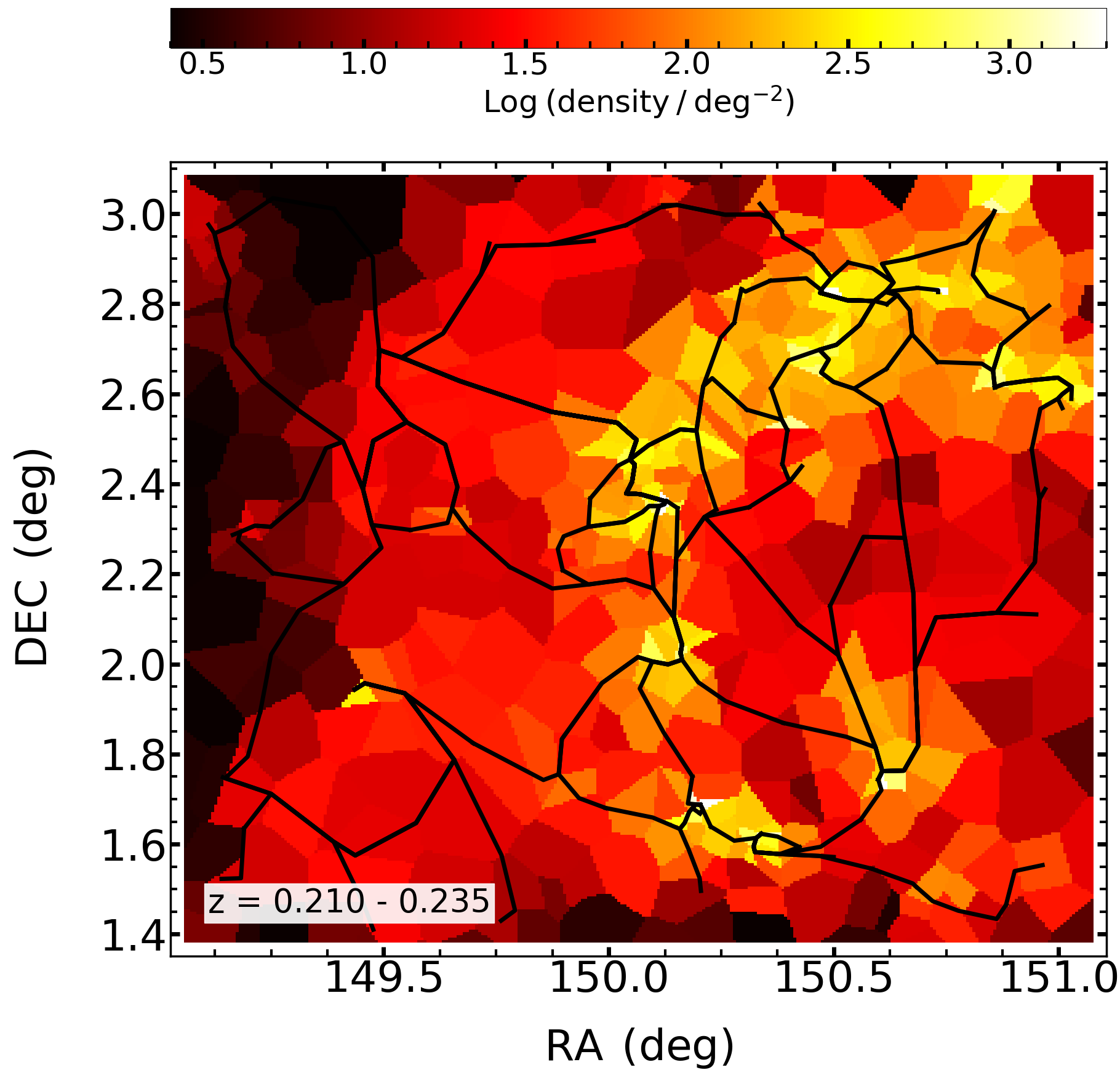}
\caption{An example {\color{black} showing a number density map} in the COSMOS field at $0.210<z<0.235$ created using DisPerSE. The colours indicate the local density (see colour bar), while the solid black lines show the locations of the filaments.} 
\label{fig:density_map}
\end{figure}

\begin{figure*}
\center
\includegraphics[width=\columnwidth]{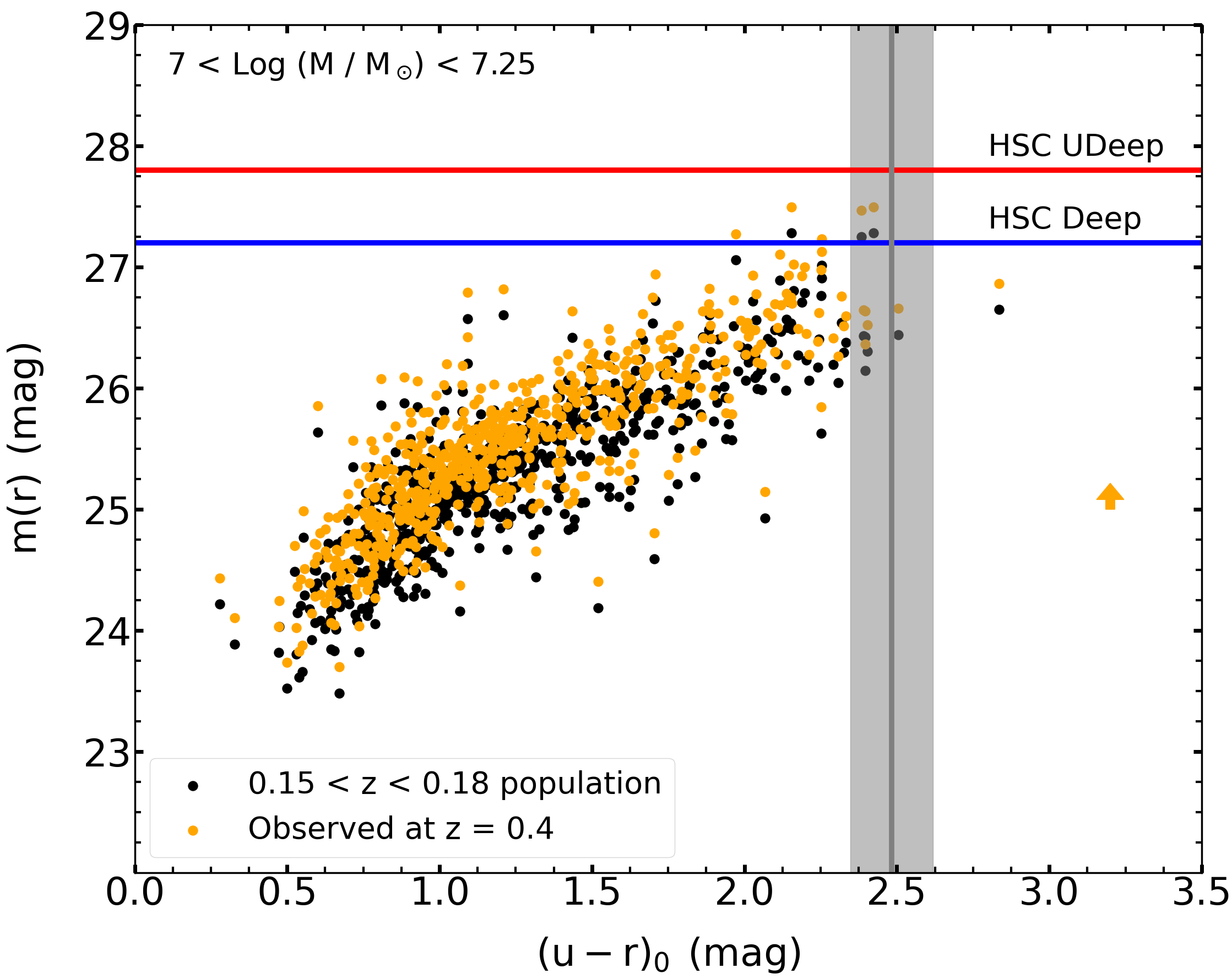}
\includegraphics[width=\columnwidth]{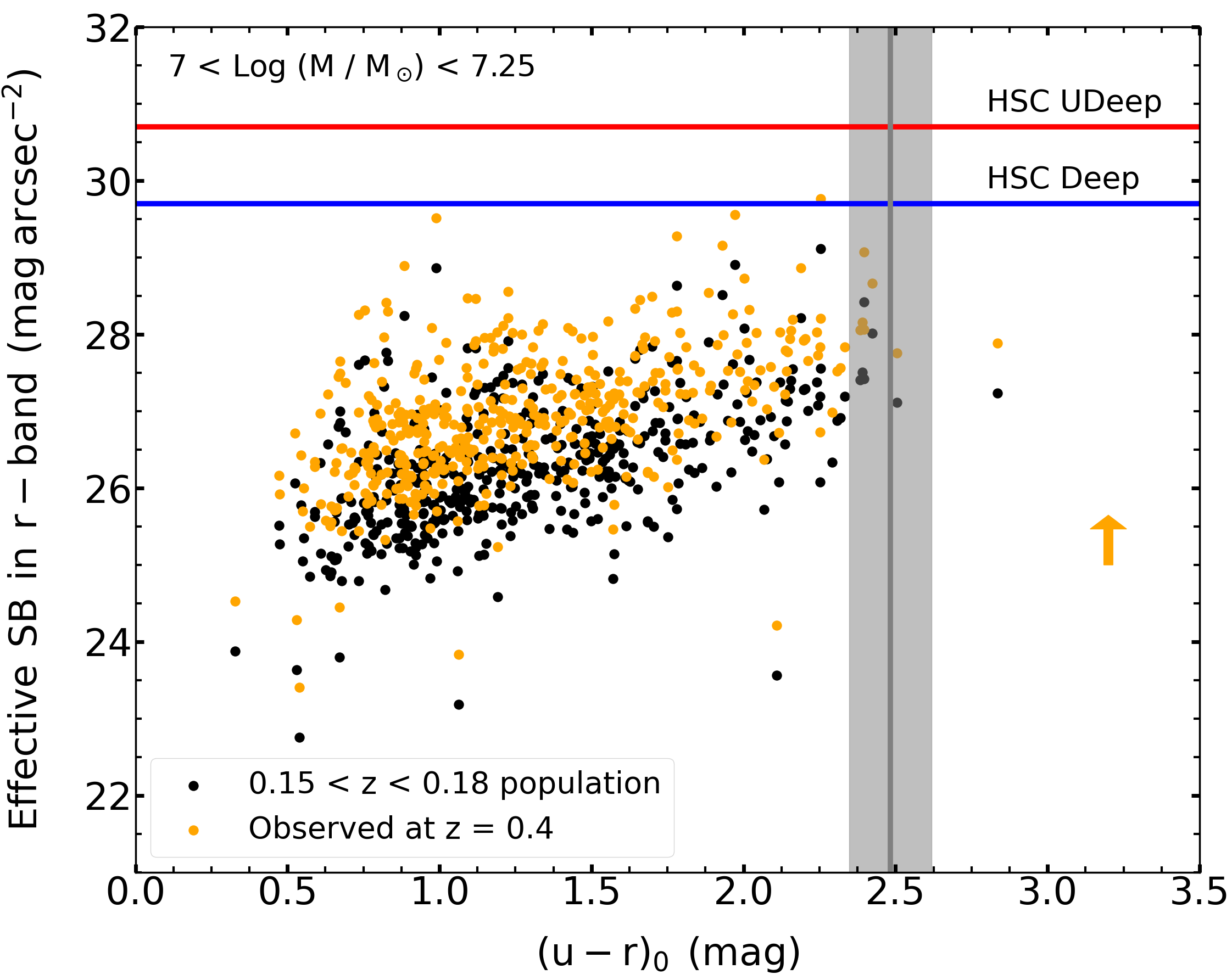}
\caption{Total magnitudes (left) and effective surface brightnesses (right) vs rest-frame $(u-r)$ colours of COSMOS2020 galaxies in the stellar mass and redshift ranges of 10$^7$ M$_{\odot}$ < $M_\star$ < 10$^{7.25}$ M$_{\odot}$ and $0.15 < z < 0.18$ respectively. The black points indicate the COSMOS2020 galaxies. The grey shaded region indicates the location of a purely old SSP that forms at $z=2$ (assuming half solar metallicity). The upper and lower limits of the grey shaded region indicate the rest-frame $(u-r)$ colour of this SSP with and without a reddening of $E(B-V)$ = 0.1 respectively The solid grey vertical line indicates the mid-point of the grey shaded region. The red and blue horizontal lines indicate the detection limits of the Deep and Ultradeep layers of the HSC-SSP imaging.}
\label{fig:completeness_thresholds}
\end{figure*}


\subsection{Cosmological hydrodynamical simulations}

{\color{black}We briefly describe the cosmological hydrodynamical simulations that are compared to the observational data in Section \ref{sec:missing_dwarfs} below. Table \ref{tab:simulation comparison} summarizes the properties of each simulation, while the sections below provide a brief outline of their key characteristics.}


\subsubsection{\textsc{NewHorizon}}

{\color{black}\textsc{NewHorizon} is a high-resolution zoom-in of a 20 Mpc diameter spherical volume of the parent Horizon-AGN simulation \citep{Dubois2014,Kaviraj2017}, which uses the adaptive mesh refinement code \textsc{RAMSES} \citep{Teyssier2002}. \textsc{NewHorizon} achieves a maximum spatial refinement of 34 pc and a stellar mass resolution of $1.3\times10^4$ M$_\odot$ \citep{Dubois2021}. It is not calibrated to match specific observational data, except for AGN feedback, whose efficiency is calibrated using the local $M_{\rm BH}-M_{*}$ relation. Star formation is based on dense, self-gravitating gas, with an efficiency regulated by local turbulence. Feedback is implemented via mechanical supernova feedback from Type II supernovae \citep{Kimm2014} and the interstellar medium (ISM) is partially resolved, allowing for multi-phase structure to emerge.}


\subsubsection{\textsc{TNG50}}

\textsc{TNG50} is part of the IllustrisTNG suite and simulates a 50 Mpc cosmological box using the moving mesh code \textsc{AREPO} \citep{Springel2010}. It includes magneto-hydrodynamics and is explicitly calibrated to match a range of low-redshift observables, such as the stellar mass function and galaxy sizes in the intermediate and high-mass regime. Star formation follows a Kennicutt-Schmidt relation \citep[][]{Kennicutt1998} within an idealized two-phase ISM governed by an effective equation of state \citep{Springel2003}. Feedback from AGN, Type I and Type II supernovae, and stellar winds is implemented through a combination of thermal and kinetic modes. Stellar feedback is injected as kinetic winds and initially decoupled from the dense ISM. AGN feedback includes both thermal energy injection in the high-accretion regime and decoupled kinetic winds in the low-accretion regime.


\subsubsection{\textsc{FIREBox}}

{\color{black}\textsc{FIREBox} simulates a 22 Mpc cosmological box based on the \textsc{FIRE-2} model, which is run using the mesh-free finite mass code \textsc{GIZMO} \citep{Hopkins2015} and provides high spatial (20 pc) and mass ($m_\star=6\times10^4$ M$_\odot$) resolution. The code is designed to minimise the number of tunable parameters, eschewing empirical calibration in favour of a physics-based approach that aims to model well-understood physical processes at sufficient resolution. Star formation occurs in explicitly cold, dense, self-gravitating gas with an assumed 100 per cent efficiency per local freefall time \citep{Hopkins2018}. Feedback includes mechanical energy and momentum from both Type I and Type II supernovae, stellar winds and radiation, but omits AGN feedback. The ISM is partially resolved, allowing for multiphase gas dynamics.} 

\begin{table*}
\caption{Key characteristics of the cosmological hydrodynamical simulations used in this study.}
\label{tab:simulation comparison}
\begin{tabular}{@{}llll@{}}
\toprule
                   & \textsc{NewHorizon} \citep{Dubois2021}                                                   & TNG50 \citep{Nelson2019,Pillepich2019}                                                       & \textsc{FIREbox} \citep{Feldmann2023} \\ \midrule
Code               & RAMSES \citep{Teyssier2002}                                                                                          & AREPO  \citep{Springel2010}                                                                                                       & \textsc{gizmo} \citep{{Hopkins2015}}        \\
Volume             & Zoom of 20 Mpc spherical region                                                             & 50 Mpc box                                                                                                    & 22 Mpc box         \\
Spatial resolution & 34 pc                                                                                       & 100 -- 140 pc                                                                                               & 20 pc         \\
Mass resolution    & $m_{\star}=1.3\times10^{4}$ M$_{\odot}$                                                            & $m_{\star}=8.5\times10^{4}$ M$_{\odot}$                                                                         & $m_{\star}=6\times10^{4}$ M$_{\odot}$          \\
Environment        & Maximum halo mass $\sim 10^{13}$ M$_{\odot}$                                                                & Maximum halo mass $\sim 10^{14}$ M$_{\odot}$                                                                             & Maximum halo mass $\sim 10^{13}$ M$_{\odot}$         \\
Star formation     & \begin{tabular}[c]{@{}l@{}}\hspace{-11pt} Dense, self-gravitating gas;\\ \hspace{-9pt}efficiency modulated by turbulence\end{tabular}                                                                           & \hspace{-10pt}\begin{tabular}[c]{@{}l@{}}Schmidt law in two-phase ISM;\\ density threshold-based\\ \citep{Springel2003}\end{tabular}                                                                                                & \begin{tabular}[c]{@{}l@{}}\hspace{-10pt}Cold, dense, self-gravitating gas;\\\hspace{-10pt}100\% efficiency per freefall time\\\hspace{-10pt}\citep{Hopkins2018}\end{tabular} \\
SN feedback        & \begin{tabular}[c]{@{}l@{}}\hspace{-10pt}Mechanical feedback from SN\\\hspace{-10pt}Type II \citep{Kimm2014}\end{tabular} & \begin{tabular}[c]{@{}l@{}}\hspace{-10pt}Direct heating and delayed kinetic winds\\\hspace{-10pt}from SNe I/II and stellar winds\\\hspace{-10pt}\citep{Springel2003}\end{tabular} &  \begin{tabular}[c]{@{}l@{}}\hspace{-10pt}Mechanical feedback from SNe I/II, stellar\\\hspace{-10pt}winds and radiation \citep{Hopkins2018},\\\hspace{-10pt}no AGN feedback\end{tabular}     \\
ISM physics        & Partially resolved multiphase ISM                                                                & \begin{tabular}[c]{@{}l@{}}\hspace{-10pt}Idealised two-phase model with\\\hspace{-10pt}effective equation of state\end{tabular}                                                                                & Partially resolved multiphase ISM         \\ \bottomrule
\end{tabular}
\end{table*}


\section{Completeness}
\label{sec:completeness} 

We begin by quantifying the mass and redshift ranges within which mass-complete samples of galaxies can be constructed, building on our previous work. Using complete samples is desirable because it avoids the need to apply completeness corrections which carry associated uncertainties and biases \citep[see e.g.][]{Weigel2016}. In \citet{Kaviraj2025}, we have used a purely old simple stellar population (SSP), that forms at $z\sim2$, to estimate the redshifts at which galaxy populations are likely to be complete as a function of stellar mass. An SSP is defined as one where all stars form simultaneously. Assuming that this hypothetical stellar population is a faintest limiting case, it follows that, if this population is detectable in a given survey, then all galaxies will also be detectable and the galaxy population will therefore be mass-complete. 

For example, \citet{Kaviraj2025} estimate that galaxies down to 10$^8$ M$_{\odot}$ should be complete out to at least $z\sim0.4$ at the depth of the HSC data in the COSMOS field (see their Figure 4). Note that \citet{Weaver2022} come to an identical conclusion using a different technique. However, \citet{Kaviraj2025} also note that the completeness redshifts thus derived are pessimistic because the vast majority of galaxies at any stellar mass (but particularly in the dwarf regime) do not consist of purely old stellar populations (see their Figure 5). Here we refine the completeness redshifts presented in Figure 4 in \citet{Kaviraj2025} using observed dwarf galaxies from the COSMOS2020 catalogue. 

In Figure \ref{fig:completeness_thresholds} we plot the total magnitudes and effective surface brightnesses vs the rest-frame $(u-r)$ colours of galaxies in the stellar mass and redshift ranges of 10$^7$ M$_{\odot}$ < $M_\star$ 10$^{7.25}$ M$_{\odot}$ and $0.15 < z < 0.18$ respectively. We select these galaxies because \citet{Kaviraj2025} demonstrate that dwarfs with $M_\star$ $\sim$ 10$^7$ M$_{\odot}$ are complete out to at least $z\sim 0.18$. The black points indicate the COSMOS2020 galaxies. The grey shaded region indicates the location of a purely old SSP that forms at $z=2$ (assuming half solar metallicity). The lower and upper limits of the grey shaded region indicate the rest-frame $(u-r)$ colour of this SSP without any reddening and with a reddening of $E(B-V)$ = 0.1, respectively. The solid grey vertical line indicates the mid-point of the grey shaded region. The red and blue horizontal lines indicate the detection limits of the Deep and Ultradeep layers of the HSC-SSP imaging. 

As already noted by \citet{Kaviraj2025}, less than 1 per cent of dwarfs are actually consistent with a purely old SSP (i.e. reside inside the grey shaded region). It is also apparent that the vast majority of galaxies in this mass range are brighter than the detection limit of both the HSC Deep and Ultradeep imaging. The orange points show how our COSMOS2020 galaxy population moves in this parameter space if they are redshifted to $z=0.4$. The shift, indicated by the orange arrows, includes both distance dimming and a $k$-correction. Here we apply the $k$-correction from our purely old SSP that forms at $z=2$, which produces a larger $k$-correction than stellar populations with younger ages. In other words, the $k$-correction, and therefore the shift in the galaxies applied here, is likely to be maximal. 

We find that, while \citet{Kaviraj2025} show that the galaxy population at $M_\star$ $\sim$ 10$^7$ M$_{\odot}$ is complete to \textit{at least} $z\sim 0.18$, in reality, since hardly any galaxies are consistent with purely old stellar populations, galaxies with $M_\star$ $\sim$ 10$^7$ M$_{\odot}$ are likely to remain complete out to around $z\sim0.4$ in both the Deep and Ultradeep layers of the HSC imaging. In these mass and redshift ranges complete samples of galaxies can be constructed, removing the need to apply completeness corrections. 


\section{Measurement of mass function uncertainties}
\label{sec:mass_function_uncertainties}

The uncertainty associated with the galaxy stellar mass function ($\sigma_\Phi$) is estimated by adding, in quadrature, the Poisson noise ($\sigma_P$), the uncertainties on the stellar masses that are calculated via SED fitting ($\sigma_{M})$ and cosmic variance ($\sigma_{CV}$), such that $\sigma_\Phi$$=$$\sqrt{\rm\sigma_P^2+\sigma_{M}^2+\sigma_{CV}^2}$. We briefly describe each contribution to the uncertainty below. 


\subsection{Poisson noise}

The Poisson noise measures the uncertainty in the number counts (for low number counts the Poisson noise can make a significant contribution to the overall uncertainty in the mass function). We define the Poisson uncertainty as $\sigma_{\rm P} = \sqrt{\rm N}/V$ where $\rm N$ represents the number count in a given mass bin and $V$ is the maximum volume within which objects in that mass bin can be detected, given the redshift range and mass completeness of the sample.


\subsection{Uncertainty on the stellar mass estimated via SED fitting}
\label{sec:MonteCarlo}

We calculate $\sigma_{\rm M}$ via a Monte Carlo approach, in which 1000 independent realisations of the galaxy stellar mass function are constructed, by drawing a random stellar mass value from the stellar mass likelihood distribution of each object in each realization. We assume a Gaussian likelihood distribution with a standard deviation that is equal to the stellar mass uncertainty measured by the SED fitting (in log stellar mass space). We estimate $\sigma_{\rm M}$ by calculating the standard deviation of $\Phi$} in each stellar mass bin across all realizations\footnote{ It is worth noting here that the systematic statistical bias induced by more low mass galaxies scattering to high masses than vice versa, due to the intrinsic shape of the mass function (Eddington bias) \citep[e.g.][]{Obreschkow2018}, is negligible in this work. Appendix \ref{app:eddington_bias} demonstrates this by showing that the mean galaxy stellar mass functions obtained from the perturbed stellar masses are very similar (within $\sim$0.01 dex) of the unperturbed mass functions in both the COSMOS and XMM-LSS fields.}.




\subsection{Cosmic variance}

Cosmic variance refers to the fact that the statistical properties of galaxies (e.g. the distribution of stellar masses) are affected by LSS. This can lead to field-to-field differences in different parts of the sky, which are larger than the Poisson noise. Larger sky areas lead to smaller cosmic variance and vice-versa. We estimate the uncertainties due to cosmic variance using the Python version of the Cosmic Variance Cookbook \citep{Jespersen2025}, originally developed in IDL by \citet[][]{Moster2010}.  


\section{The dwarf stellar mass function in different environments}
\label{sec:mass_functions}


It has been shown in a variety of studies, using both simulations and observations, that galaxy properties such as SFR, colour and stellar mass, at least in the massive-galaxy regime, change as a function of distance from LSS traced by the nodes and filaments of the cosmic web. For example, galaxies that are more massive and more quiescent typically reside closer to nodes and filaments, while environments further away from LSS typically host lower mass galaxies with higher levels of star formation. This dependence on the distance from LSS has been demonstrated both in observational \citep[e.g.][]{Laigle2018} and theoretical \citep[e.g.][]{Kraljic2018,Hasan2023} work, at least in the low and intermediate redshift Universe ($z<2$). Indeed, galaxy properties may change more noticeably with varying distance to LSS than with just the numerical value of local density \citep[e.g.][]{Laigle2018}. Recent empirical work using the SDSS \citep{Zarattini2025} suggests that galaxy SFRs begin to show statistical differences, compared to those in the field, {\color{black} for galaxies located within 5 Mpc of filaments}. Galaxy colours show similar changes {\color{black} for galaxies located within $\sim$1 Mpc of filaments}. This suggests that distance from LSS, on scales of around 5 Mpc or less, plays an important role in influencing the physical properties of galaxies, at least in the massive-galaxy regime.

In light of these results we use the distance from LSS as our metric for environment. We note first that, while we explore the COSMOS dataset down to stellar masses of 10$^7$ M$_\odot$, we restrict the XMM-LSS dataset to stellar masses above 10$^8$ M$_\odot$. This is because the mass function in XMM-LSS shows an unexpected downturn at $M_\star$ < 10$^8$ M$_\odot$, which is not shared by its COSMOS counterpart (see the discussion in Appendix \ref{app:medium_band_filters} and Figure \ref{fig:COSMOS_XMM_comp}). We speculate that this inconsistency between XMM-LSS and COSMOS may be driven by the availability of medium-band filters in the SED fitting process, since the SED fitting in COSMOS benefits from the presence of medium-band filters while the XMM-LSS dataset does not. Indeed, when we use the same filterset as that employed for the SED fitting in XMM-LSS (i.e. broadband $ugrizyJHK$ only) on COSMOS data, we obtain a similar downturn in the COSMOS mass function (see Figure \ref{fig:COSMOS_XMM_comp}). This suggests that the lack of medium-band filters may lead to a poorer sampling of the SED which appears to affect very low mass ($M_\star$ < 10$^8$ M$_\odot$) galaxies. Note that this downturn has also been noted in previous work that has only used broadband filters for measuring SED fitted stellar masses in the XMM-LSS and COSMOS fields \citep[e.g.][]{Adams2021}. Thus, as a precaution we consider the XMM-LSS dataset only down to $M_\star$ = 10$^8$ M$_\odot$ where the XMM-LSS mass function shows good consistency with its COSMOS counterpart, regardless of the filters used in the SED fitting. This issue is discussed further in Appendix \ref{app:medium_band_filters}. 


\subsection{Variation of the mass function with distance from LSS}

\label{sec:mf_nodes_filaments}

\begin{figure}
\center
\includegraphics[width=\columnwidth]{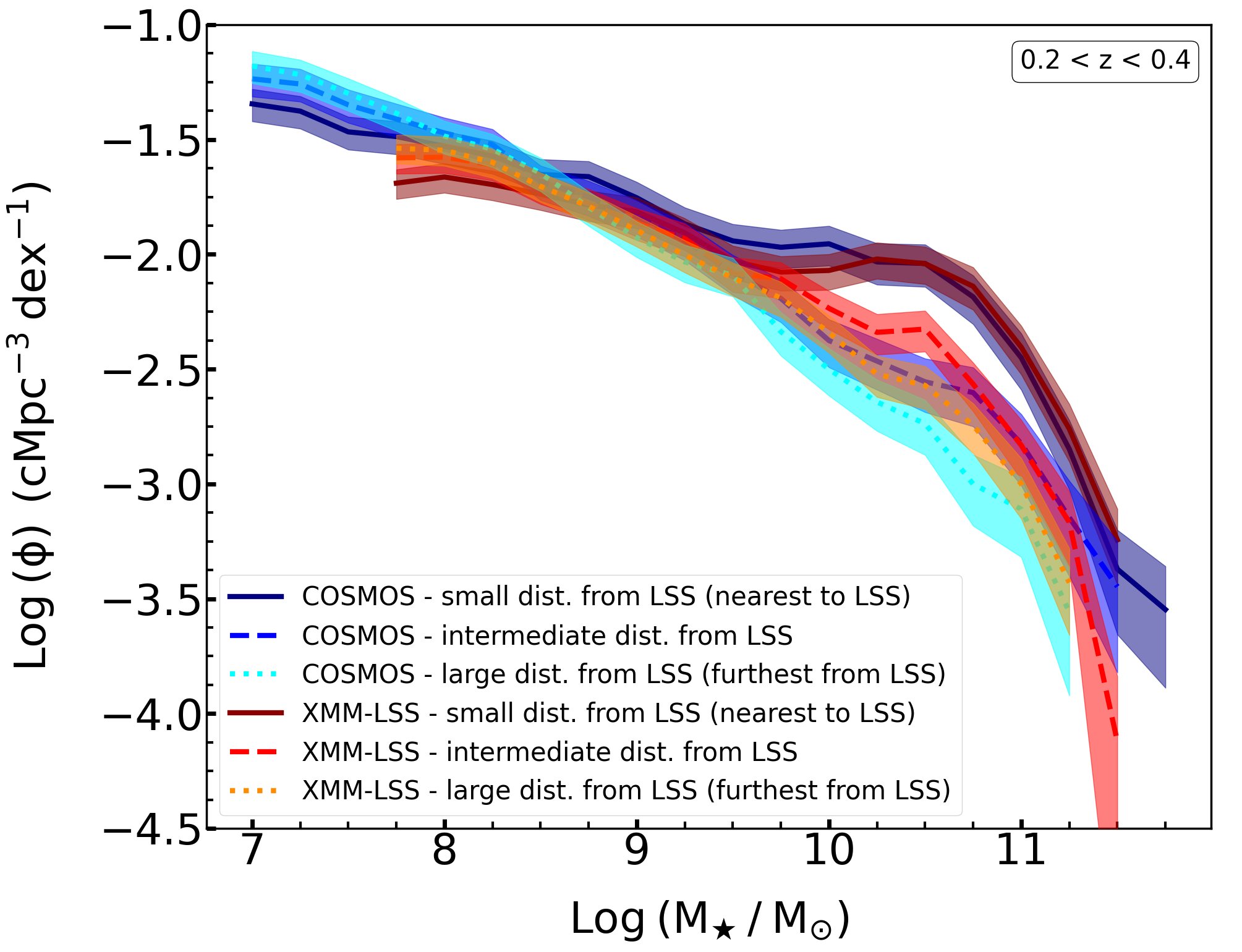}
\caption{The galaxy stellar mass function for the COSMOS and XMM-LSS fields in different environments. Galaxies at small distances from LSS (i.e. nearest to LSS) are defined as those that reside in the lowest tercile (0$^{\rm th}$ -- 33$^{\rm rd}$ percentile values) of both the distances to nodes and the distances to filaments. Galaxies at intermediate and large distances from LSS are defined as those that reside in the middle (33$^{\rm rd}$ -- 66$^{\rm th}$ percentile values) and upper (66$^{\rm th}$ -- 100$^{\rm th}$ percentile values) terciles of both the distances to nodes and the distances to filaments respectively.}  
\label{fig:MFcosmicWeb}
\end{figure}

We proceed by using the outputs from DisPerSE to explore the relationship between the galaxy stellar mass function and the proximity of galaxies to LSS. We first split our galaxies into three terciles in terms of their {\color{black} projected} distances from the nearest node and three terciles in terms of their {\color{black} projected} distances from the nearest filament. We then define three galaxy populations that reside at `small', `intermediate' and `large' distances from LSS as follows. Galaxies at small distances from LSS (i.e. nearest to LSS) are defined as those that reside in the lowest tercile (0$^{\rm th}$ -- 33$^{\rm rd}$ percentile values) of both the distances to nodes and the distances to filaments. Similarly, galaxies at intermediate and large distances from LSS are defined as those that reside in the middle (33$^{\rm rd}$ -- 66$^{\rm th}$ percentile values) and upper (66$^{\rm th}$ -- 100$^{\rm th}$ percentile values) terciles of both the distances to nodes and the distances to filaments respectively. In Figure \ref{fig:MFcosmicWeb} we show the galaxy stellar mass function for these three populations. 

Two clear differences arise when studying how the mass function behaves with respect to distance from LSS, regardless of the field (i.e. COSMOS or XMM-LSS) being considered. First, the knee of the mass function becomes more pronounced in regions that are closer to LSS. The galaxy number density around the knee of the mass function is $\sim$0.4 dex larger in galaxies that are in the lower tercile in their distance to both nodes and filaments (i.e. nearest to LSS) compared to those that are in the upper tercile (i.e. furthest from LSS). In other words, the number density of massive systems increases as we move closer to LSS. Second, the mass function in the dwarf galaxy regime at $M_\star$ $<$ 10$^{8.5}$ M$_\odot$ becomes flatter -- i.e. there are fewer dwarfs per unit volume -- as we move closer to LSS. The net effect is that the ratio of massive to dwarf galaxies increases with greater proximity to LSS. For completeness, we note that these trends also exist (although with varying significance) when considering the stellar mass function for different terciles in the distances to filaments and nodes \textit{separately} (see discussion in Appendix \ref{app:just_nodes_and_filaments} and Figure \ref{fig:MFdistanceCosmicWeb}). Figure \ref{fig:MFdistanceCosmicWeb} demonstrates that the environmental trends are driven principally by the distances of galaxies from filaments, rather than their distances from nodes. 

The trends described above are consistent with the predictions of the $\Lambda$CDM paradigm. The galaxy stellar mass distribution is predicted to depend on the surrounding large-scale environment due to biased clustering \citep{Kaiser1984}, with the non-linear collapse of over-dense regions and their location within the cosmic web determining where and how halos form \citep[e.g.][]{Bond1996}. The galaxy bias -- particularly in terms of the location of galaxies with respect to filaments \citep{Musso2018} -- is related to the distribution of dark matter, as a consequence of the large-scale tides that impact on the assembly history of halos. \citet{Kraljic2019} have used a cosmological hydrodynamical simulation to demonstrate that the overall impact of this process is two-fold. The stellar mass of galaxies increases with decreasing distance from the cosmic web (in the transverse and longitudinal directions with respect to the filament-type saddle point) while the number of massive galaxies is higher near the filaments compared to that of the lower-mass ones (see also \citealp{Song2021}), in good agreement with our empirical results.   

We continue by performing a quantitative analysis of the differences that are observed in the mass function in different environments. Galaxy mass functions have often been parametrised using a Schechter function \citep{Schechter1976}. An extension of this procedure is to use a double Schechter function, which provides a better representation of the observed upturn in galaxy number densities when low mass galaxies are included, at least at low and intermediate redshift \citep[e.g.][]{Baldry2012,Weaver2022}. The double Schechter function is expressed as follows:

\begin{equation}
\begin{split}
\rm\Phi \: \textit{d} \: log \: \textit{M} = ln(10) \: e^{-10^{log \: \textit{M} - log \: \textit{M}^\star}} \:\:\:\:\:\:\:\:\:\:\:\:\:\:\:\:\:\:\:\:\:\:\:\, \\  \times  [ \Phi^\star_1 (10^{\rm log \: \textit{M} - log \: \textit{M}^\star})^{\alpha_1+1} \:\:\:\:\:\:\:\:\:\:\:\:\:\:\:\:\, \\ +  \:\: \Phi^\star_2 (10^{\rm log \: \textit{M} - log \: \textit{M}^\star})^{\alpha_2+1} ] \: \textit{d} \: \rm log \: \textit{M}
\end{split}
\end{equation}\\

Here $M^*$ represents a characteristic mass scale (which corresponds to the knee of the mass function) beyond which there is an exponential decline in galaxy number densities at high stellar masses. $\alpha_1$ and $\alpha_2$ are the slopes of the two Schechter functions which have normalizations $\Phi^\star_1$ and $\Phi^\star_2$ respectively. The slope of the fitted function at stellar masses much lower than the knee of the mass function is controlled by the more negative $\alpha$ value, while the slope nearer the knee is controlled by the more positive $\alpha$ value. 

\begin{figure}
\center
\includegraphics[width=\columnwidth]{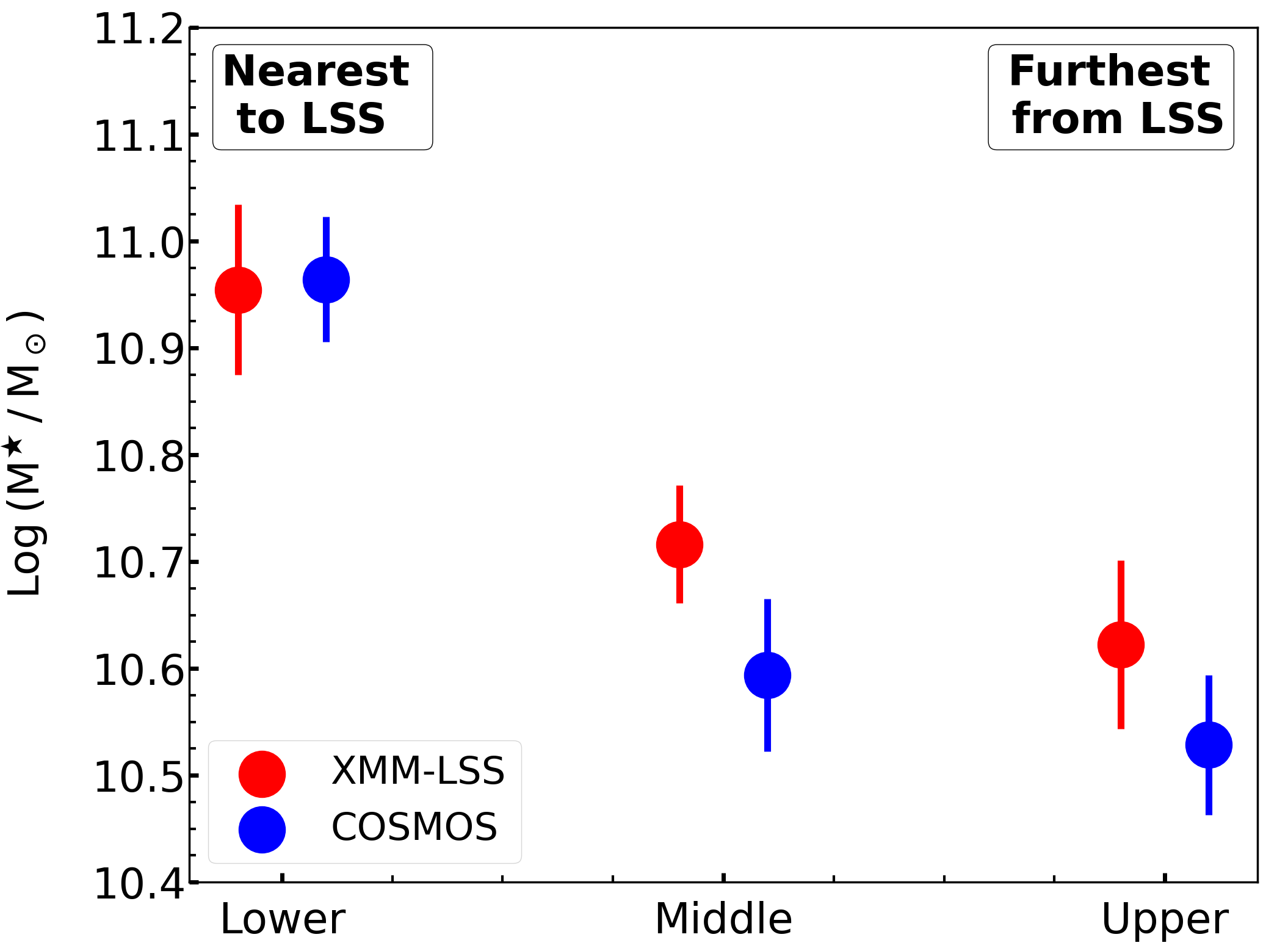}
\includegraphics[width=0.995\columnwidth]{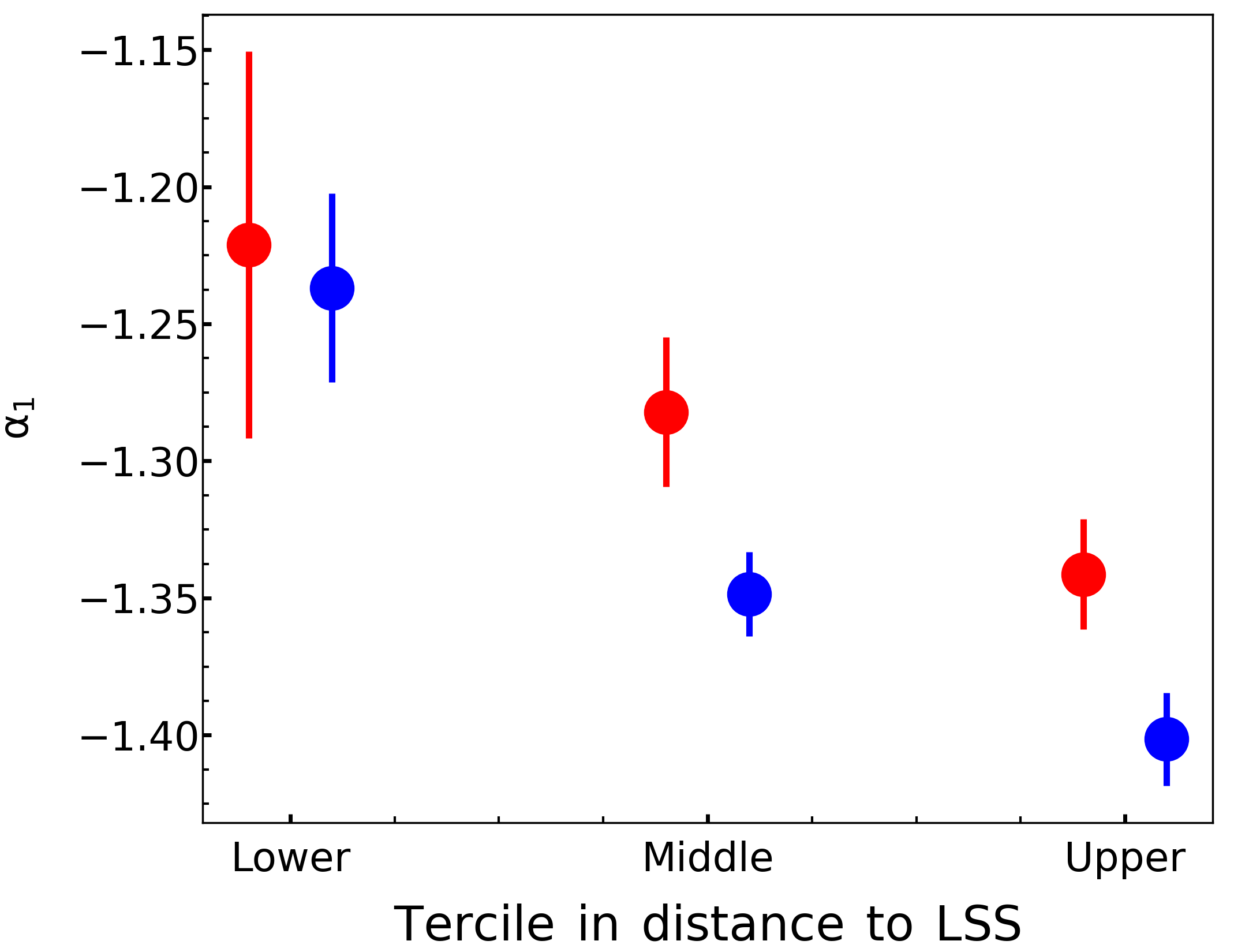}
\caption{Best-fit parameter values and their uncertainties for M$^\star$ and $\alpha_1$ from the double Schechter fits to the mass functions in the COSMOS and XMM-LSS fields in different environments. Galaxies at small distances from LSS (i.e. nearest to LSS) are defined as those that reside in the lowest tercile (0$^{\rm th}$ -- 33$^{\rm rd}$ percentile values) of both the distances to nodes and the distances to filaments. Galaxies at intermediate and large distances from LSS are defined as those that reside in the middle (33$^{\rm rd}$ -- 66$^{\rm th}$ percentile values) and upper (66$^{\rm th}$ -- 100$^{\rm th}$ percentile values) terciles of both the distances to nodes and the distances to filaments respectively.} 
\label{fig:schparams}
\end{figure}

We fit this double Schechter function to our mass functions in different environments. In Figure \ref{fig:schparams} we present the values of $M^*$ and $\alpha_1$ (the more negative of the $\alpha$ values, which represents the slope of the mass function in the dwarf regime) for the three galaxy populations defined above. Depending on the field in question, the values of $M^*$ and $\alpha_1$ become more positive by 0.3 -- 0.4 dex and 0.1 -- 0.2 dex, respectively, in galaxies that are in the lower tercile in their distances to both nodes and filaments (i.e. nearest to LSS), compared to those that are in the lower tercile (i.e. furthest from LSS). In other words, the knee of the mass function moves to larger stellar masses and the slope in the dwarf regime becomes flatter as we move closer to LSS (as is evident from visual inspection of Figure \ref{fig:MFcosmicWeb}). The best-fit parameters of our Schechter fits are presented in Table \ref{tab:schechter_params}. 

\subsection{Comparison with the observational literature}

{\color{black} We complete this section by comparing our results to the recent observational literature. It is worth noting first that complete consensus about the role of environment on the shape of the mass function, especially in the dwarf regime, does not yet exist. This is likely to be due -- at least partly -- to the heterogeneity of methods that are employed to define environment in the literature and cosmic variance. Nevertheless, our results are in good agreement with some recent studies that also use deep photometric data to extract mass functions (typically via SED fitting) and probe its evolution as a function of environment. 

For example, \citet{Malavasi2017} find that, in the UltraVISTA-COSMOS field in our redshift range of interest, denser environments exhibit a larger number density of massive galaxies and a smaller number density of dwarfs. A very similar result is found by \citet{Etherington2017}, using a statistical sample of $\sim$3.2 million galaxies observed as part of the Dark Energy Survey (DES). They find that the number density of dwarfs in denser environments is lower than that in more sparse environments by $\sim$0.4 dex and that massive galaxies in denser environments outnumber their counterparts in sparse environments by a similar amount. The results of both these studies are very similar to the results of this study, shown in Figure \ref{fig:MFcosmicWeb}. Similarly, \citet{Mortlock2015} use deep HST observations in the GOODS and UDS fields to show that more massive galaxies exist per unit volume in denser environments at our redshifts of interest (but interestingly do not report significant differences in the dwarf regime in different environments). 

All three studies described above define environment by calculating galaxy number counts within cylinders or conical frusta, where the projected radii of these shapes is around 1 Mpc or less, indicating that they trace local environment on relatively small spatial scales. As a comparison we show, in Figure \ref{fig:phys_scale}, the distribution of distances from the nearest filaments for the galaxy population studied here. The median value of this distribution, shown using the vertical blue line, lies close to 1 Mpc while the distances reach a maximum value of $\sim$6 Mpc. The spatial scales probed here are therefore similar to those in the studies described above, which explains the consistency between our results. 

Together with the literature, our results suggest that distance from LSS has a strong influence on galaxy stellar mass, on physical scales of around $\sim$ 1 Mpc or less. Interestingly, both \citet{Malavasi2017} and \citet{Zarattini2025}, as discussed above, suggest that these effects start to diminish once physical scales larger than around 4 -- 5 Mpc are probed. 

\begin{figure}
\center
\includegraphics[width=\columnwidth]{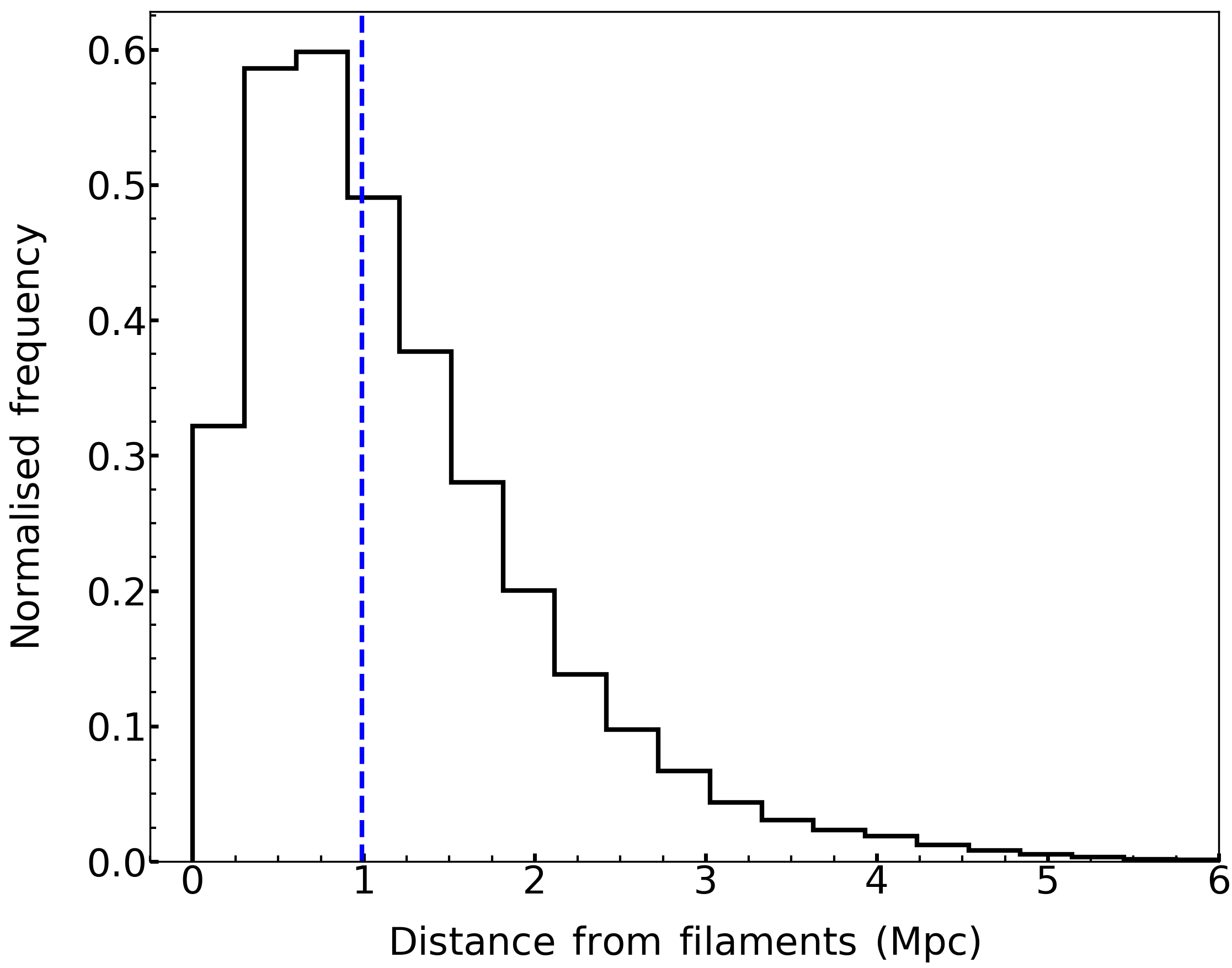}
\caption{The distribution of projected distances from the nearest filaments for galaxies in the COSMOS and XMM-LSS fields. The median distance is shown using the dashed blue vertical line.}
\label{fig:phys_scale}
\end{figure}


\section{Is there a missing dwarfs problem in the context of the 
$\Lambda$CDM paradigm?}
\label{sec:missing_dwarfs} 

As noted in the introduction, the missing satellites problem is a longstanding potential tension in the $\Lambda$CDM paradigm between the predicted and observed number densities of low mass galaxies. Specifically, it has been claimed that observational data may show a deficit of dwarfs compared to the predictions of simulations, which makes the observed stellar mass function of low-mass galaxies in our local neighbourhood much flatter than its theoretical counterpart \citep[e.g.][]{Moore1999,Klypin1999}. However, detailed explorations of this issue are largely restricted to dwarfs around either the Milky Way or other local massive galaxies \citep[e.g.][and references therein]{Tanaka2018,Smercina2018,Bennet2020,Nashimoto2022}. 

A potential solution to the missing satellites problem involves invoking warm or fuzzy dark matter, which suppresses the formation of low-mass halos and therefore the number density of dwarf galaxies \citep[e.g.][]{Hu2000,Polisensky2011,Lovell2012,Marsh2016}. Alternatively, the inclusion of baryonic physics can also reduce the number density of dwarf galaxies within the framework of the $\Lambda$CDM paradigm, without altering the properties of dark matter, largely through the reduction of galaxy SFRs via star formation \citep[e.g.][]{Geen2013,Hazenfratz2024} and AGN \citep[e.g.][]{Dashyan2018,Kaviraj2019,Koudmani2021,Davis2022} feedback and ionising UV radiation \citep[e.g.][]{Forbes2016,Emerick2019}. Recent work using high-resolution hydrodynamical simulations suggests that, when such realistic baryonic physics is implemented, the number densities of low-mass galaxies around Milky Way like galaxies in $\Lambda$CDM-based simulations can be consistent with their observed counterparts \citep[e.g.][]{Sawala2016,Santos-Santos2022,Jeon2025}. 

It is worth noting that dwarfs in close proximity to the Milky Way are, by construction, satellites. This is why past explorations of this issue have been crafted in terms of missing \textit{satellite} galaxies. However, if such a problem does indeed exist, it cannot be restricted to just the vicinity of massive galaxies like the Milky Way. In other words, a discrepancy between the predicted and observed number densities should manifest itself as a missing \textit{dwarfs} problem in the context of the $\Lambda$CDM paradigm. The deep data now available provides an unprecedented opportunity to explore this issue in the general dwarf population across all environments selected from a deep-wide survey, rather than only the relatively high-density neighbourhoods of very nearby massive Milky Way like galaxies. Note, however, that in our case, this exercise is restricted to $M_{\star}$ > 10$^7$ M$_{\odot}$ and $z<0.4$, where we can construct complete galaxy samples. 

We complete our study by exploring whether a comparison between the observed and theoretical stellar mass functions suggests the presence of a missing dwarfs problem in the context of $\Lambda$CDM. In Figure \ref{fig:mf_general}, we compare the observed mass functions in the COSMOS and XMM-LSS fields, and an ensemble of empirical results drawn from the literature, to the predicted mass functions from three cosmological hydrodynamical simulations (\textsc{NewHorizon}, TNG50 and \textsc{FIREbox}). For this exercise, we study the total mass function, constructed from all galaxies in both the observations and the simulations in our mass and redshift ranges of interest. There is excellent agreement between the observed mass functions calculated in this study and their empirical counterparts from the literature. 

We note that the mass functions from the three simulations do not completely agree with each other. While there is good agreement at the low mass end of the galaxy population considered in this study ($M_{\star}$ < 10$^8$ M$_{\odot}$), the disagreement between the different theoretical mass functions can be as large as $\sim$0.5 dex at some points within the dwarf regime. The survivability and evolution of galaxies at dwarf masses can depend on simulation resolution, particularly in cluster environments \citep{Martin2024}. But, given the low-density environments probed by these simulations and the similar mass/spatial resolutions and cosmology adopted by each model, these discrepancies are more likely to be driven by differences in the implementation of baryonic physics, such as baryonic feedback and ISM gas physics \citep[e.g.][]{Martin2025}. 

It is also worth noting that the recent literature suggests that stellar masses derived from SED fitting systematically underestimates the true stellar masses of galaxies \citep[e.g.][]{Laigle2019,Delosreyes2024}. In particular, \citet{Laigle2019} demonstrate that performing SED fitting using the same filters as those used in this study underestimates the stellar mass by $\sim$0.12 dex at $M_{\star}$ $\sim$ 10$^9$ M$_{\odot}$, with the discrepancy increasing with decreasing stellar mass. Extrapolation of the results of their study suggests that the underestimate could be $\sim$0.25 dex at $M_{\star}$ $\sim$ 10$^7$ M$_{\odot}$, with an average value of $\sim$0.2 dex in the dwarf regime. While Figure \ref{fig:mf_general} already demonstrates reasonable consistency between theory and data, raising the observed mass function by $\sim$0.2 dex improves the consistency between the observed and theoretical mass functions even further. To illustrate this point we show, in Figure \ref{fig:mf_appendix_general}, a version of Figure \ref{fig:mf_general} where the observed mass functions in the dwarf regime have been {\color{black} shifted} by 0.2 dex. 

Given these two points, we do not find evidence that there is a missing dwarfs problem in the observational data, in line with recent work that suggests that the missing satellites problem is also solved when baryonic physics is employed in a realistic manner in simulations. Nevertheless, further work is desirable to fully explore this issue in low density environments, using both higher resolution simulations in larger volumes, in order to probe the dwarf regime down to lower stellar masses, and data from forthcoming surveys like LSST \citep{Ivezic2019} which offer large footprints that can alleviate the effect of cosmic variance. 

\begin{figure}
\center
\includegraphics[width=\columnwidth]{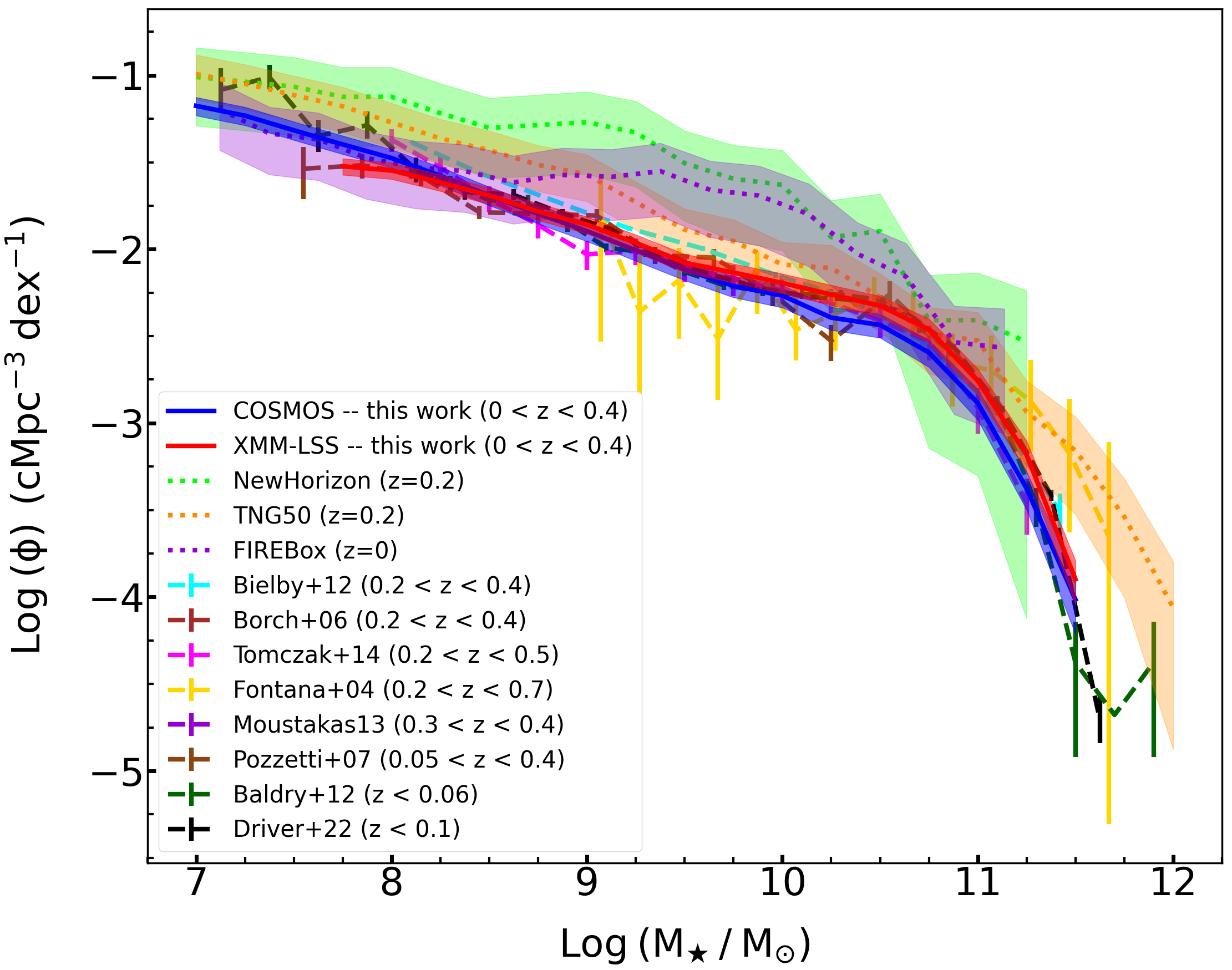}
\caption{The observed galaxy stellar mass function in the nearby Universe in the COSMOS and XMM-LSS fields and an ensemble of observations from the literature, compared to theoretical mass functions from three high resolution cosmological hydrodynamical simulations (\textsc{NewHorizon}, TNG50 and \textsc{FIREbox}). The mass functions from the literature are taken from \citet{Driver2022}, \citet{Baldry2012}, \citet{Bielby2012}, \citet{Borch2006}, \citet{Tomczak2014}, \citet{Fontana2004}, \citet{Moustakas2013} and \citet{Pozzetti2007}.}
\label{fig:mf_general}
\end{figure}


\section{Summary}
\label{sec:summary} 

We have combined deep photometric data in the COSMOS and XMM-LSS fields with high resolution cosmological hydrodynamical simulations to explore two key open questions in our understanding of dwarf-galaxy evolution: (1) how does the stellar mass function, particularly in the dwarf regime, vary with environment, defined here as the distance from LSS? (2) is there a generic `missing dwarfs' problem in the context of the $\Lambda$CDM paradigm, when all environments -- and not just the regions around Milky Way like galaxies -- are considered? The unprecedented depth of the observational survey data now available in COSMOS and XMM-LSS facilitates the construction of mass-complete galaxy populations, down to $M_{\star}$ $\sim$ 10$^7$ M$_{\odot}$ and out to $z\sim0.4$. This enables us to tackle these questions in an unbiased manner, without the need for completeness corrections. Our main conclusions are as follows:

\begin{itemize}

    \item Distance from LSS has a strong influence on the shape of the galaxy stellar mass function at both the dwarf and massive-galaxy ends. As we move closer to LSS, the number densities of massive and dwarf galaxies increase and decrease respectively, resulting in a higher ratio of massive to dwarf galaxies. 

    \item Depending on the field in question (i.e. COSMOS or XMM-LSS), the values of $M^*$ and $\alpha_1$ become more positive by 0.3 -- 0.4 dex and 0.1 -- 0.2 dex, respectively, in galaxies that are nearest to LSS (i.e. in the lower tercile in their distances to both nodes and filaments) compared to those that are furthest from LSS (i.e. in the upper tercile in their distances to both nodes and filaments). In other words, the knee of the mass function moves to larger stellar masses and the slope in the dwarf regime becomes flatter as we move closer to LSS. 
    
    \item While there is reasonable agreement between the predicted stellar mass functions of the TNG50, FIREBox and NewHorizon simulations in the stellar mass range $M_{\star}$ < 10$^8$ M$_{\odot}$, the disagreement between the different theoretical mass functions can be as large as $\sim$0.5 dex at some points within the dwarf regime (e.g. just below the knee). These discrepancies are likely to be driven by differences in the implementation of baryonic physics in the different simulations.  
    
    \item A comparison between the observed and theoretical stellar mass functions suggests that there is no generic missing dwarfs problem in $\Lambda$CDM, when all environments -- and not just the regions around Milky Way like galaxies -- are considered. This is a broader, generalized version of recent results which have shown that there is no missing satellites problem in $\Lambda$CDM when satellite populations around massive galaxies like the Milky Way are considered. 

\end{itemize}


\section*{Acknowledgements}

We warmly thank the anonymous referee for many constructive suggestions that helped us improve the quality of the original manuscript. SK, IL and AEW acknowledge support from the STFC (grant numbers ST/Y001257/1 and ST/X001318/1). SK also acknowledges a Senior Research Fellowship from Worcester College Oxford. S.K.Y. acknowledges support from the Korean National Research Foundation (RS-2025-00514475; RS-2022-NR070872).

This research has made use of the Horizon cluster on which the \textsc{NewHorizon} simulation was post-processed, hosted by the Institut d'Astrophysique de Paris. We warmly thank S.~Rouberol for running it smoothly. This work is partially supported by the grant Segal ANR-19-CE31-0017 of the French Agence Nationale de la Recherche and by the National Science Foundation under Grant No. NSF PHY-1748958. 

This research was granted access to the HPC resources of CINES under the allocations c2016047637, A0020407637, and A0070402192 by Genci, and as a ``Grand Challenge'' project granted by GENCI on the AMD Rome extension of the Joliot Curie supercomputer at TGCC. The simulation was also supported by KISTI under KSC-2017-G2-0003, KSC-2020-CRE-0055, and KSC-2020-CRE-0280 runs. The large data transfer was supported by KREONET, which is managed and operated by KISTI.

The Hyper Suprime-Cam (HSC) collaboration includes the astronomical communities of Japan and Taiwan, and Princeton University. The HSC instrumentation and software were developed by the National Astronomical Observatory of Japan (NAOJ), the Kavli Institute for the Physics and Mathematics of the Universe (Kavli IPMU), the University of Tokyo, the High Energy Accelerator Research Organization (KEK), the Academia Sinica Institute for Astronomy and Astrophysics in Taiwan (ASIAA), and Princeton University. Funding was contributed by the FIRST program from the Japanese Cabinet Office, the Ministry of Education, Culture, Sports, Science and Technology (MEXT), the Japan Society for the Promotion of Science (JSPS), Japan Science and Technology Agency (JST), the Toray Science Foundation, NAOJ, Kavli IPMU, KEK, ASIAA, and Princeton University. This paper makes use of software developed for Vera C. Rubin Observatory. We thank the Rubin Observatory for making their code available as free software at http://pipelines.lsst.io/.

This paper is based on data collected at the Subaru Telescope and retrieved from the HSC data archive system, which is operated by the Subaru Telescope and Astronomy Data Center (ADC) at NAOJ. Data analysis was in part carried out with the cooperation of Center for Computational Astrophysics (CfCA), NAOJ. We are honored and grateful for the opportunity of observing the Universe from Maunakea, which has the cultural, historical and natural significance in Hawaii. This paper used data that is based on observations collected at the European Southern Observatory under ESO programme ID 179.A-2005 and on data products produced by CALET and
the Cambridge Astronomy Survey Unit on behalf of the UltraVISTA consortium. 


\section*{Data Availability}

{\color{black}The data used in this study are taken from \citet{Weaver2022}, \citet{Desprez2023} and \citet{Picouet2023}. The density maps were created using the DisPerSE algorithm which is described in \citet{Sousbie2011}.} 

\bibliographystyle{mnras}
\bibliography{references}

\begin{thebibliography}{}
\makeatletter
\relax
\def\mn@urlcharsother{\let\do\@makeother \do\$\do\&\do\#\do\^\do\_\do\%\do\~}
\def\mn@doi{\begingroup\mn@urlcharsother \@ifnextchar [ {\mn@doi@} {\mn@doi@[]}}
\def\mn@doi@[#1]#2{\def\@tempa{#1}\ifx\@tempa\@empty \href {http://dx.doi.org/#2} {doi:#2}\else \href {http://dx.doi.org/#2} {#1}\fi \endgroup}
\def\mn@eprint#1#2{\mn@eprint@#1:#2::\@nil}
\def\mn@eprint@arXiv#1{\href {http://arxiv.org/abs/#1} {{\tt arXiv:#1}}}
\def\mn@eprint@dblp#1{\href {http://dblp.uni-trier.de/rec/bibtex/#1.xml} {dblp:#1}}
\def\mn@eprint@#1:#2:#3:#4\@nil{\def\@tempa {#1}\def\@tempb {#2}\def\@tempc {#3}\ifx \@tempc \@empty \let \@tempc \@tempb \let \@tempb \@tempa \fi \ifx \@tempb \@empty \def\@tempb {arXiv}\fi \@ifundefined {mn@eprint@\@tempb}{\@tempb:\@tempc}{\expandafter \expandafter \csname mn@eprint@\@tempb\endcsname \expandafter{\@tempc}}}

\bibitem[\protect\citeauthoryear{{Adams}, {Bowler}, {Jarvis}, {H{\"a}u{\ss}ler}  \& {Lagos}}{{Adams} et~al.}{2021}]{Adams2021}
{Adams} N.~J.,  {Bowler} R.~A.~A.,  {Jarvis} M.~J.,  {H{\"a}u{\ss}ler} B.,   {Lagos} C.~D.~P.,  2021, \mn@doi [\mnras] {10.1093/mnras/stab1956}, \href {https://ui.adsabs.harvard.edu/abs/2021MNRAS.506.4933A} {506, 4933}

\bibitem[\protect\citeauthoryear{{Aguerri}, {Iglesias-P{\'a}ramo}, {V{\'\i}lchez}, {Mu{\~n}oz-Tu{\~n}{\'o}n}  \& {S{\'a}nchez-Janssen}}{{Aguerri} et~al.}{2005}]{Aguerri2005}
{Aguerri} J.~A.~L.,  {Iglesias-P{\'a}ramo} J.,  {V{\'\i}lchez} J.~M.,  {Mu{\~n}oz-Tu{\~n}{\'o}n} C.,   {S{\'a}nchez-Janssen} R.,  2005, \mn@doi [\aj] {10.1086/431360}, \href {https://ui.adsabs.harvard.edu/abs/2005AJ....130..475A} {130, 475}

\bibitem[\protect\citeauthoryear{{Aihara} et~al.,}{{Aihara} et~al.}{2019}]{Aihara2019}
{Aihara} H.,  et~al., 2019, \mn@doi [\pasj] {10.1093/pasj/psz103}, \href {https://ui.adsabs.harvard.edu/abs/2019PASJ...71..114A} {71, 114}

\bibitem[\protect\citeauthoryear{{Alam} et~al.,}{{Alam} et~al.}{2015}]{Alam2015}
{Alam} S.,  et~al., 2015, \mn@doi [\apjs] {10.1088/0067-0049/219/1/12}, \href {https://ui.adsabs.harvard.edu/abs/2015ApJS..219...12A} {219, 12}

\bibitem[\protect\citeauthoryear{{Arnouts} et~al.,}{{Arnouts} et~al.}{2002}]{Arnouts2002}
{Arnouts} S.,  et~al., 2002, \mn@doi [\mnras] {10.1046/j.1365-8711.2002.04988.x}, \href {https://ui.adsabs.harvard.edu/abs/2002MNRAS.329..355A} {329, 355}

\bibitem[\protect\citeauthoryear{{Ashby} et~al.,}{{Ashby} et~al.}{2013}]{Ashby2013}
{Ashby} M.~L.~N.,  et~al., 2013, \mn@doi [\apj] {10.1088/0004-637X/769/1/80}, \href {https://ui.adsabs.harvard.edu/abs/2013ApJ...769...80A} {769, 80}

\bibitem[\protect\citeauthoryear{{Ashby} et~al.,}{{Ashby} et~al.}{2015}]{Ashby2015}
{Ashby} M.~L.~N.,  et~al., 2015, \mn@doi [\apjs] {10.1088/0067-0049/218/2/33}, \href {https://ui.adsabs.harvard.edu/abs/2015ApJS..218...33A} {218, 33}

\bibitem[\protect\citeauthoryear{{Ashby} et~al.,}{{Ashby} et~al.}{2018}]{Ashby2018}
{Ashby} M.~L.~N.,  et~al., 2018, \mn@doi [\apjs] {10.3847/1538-4365/aad4fb}, \href {https://ui.adsabs.harvard.edu/abs/2018ApJS..237...39A} {237, 39}

\bibitem[\protect\citeauthoryear{{Atek} et~al.,}{{Atek} et~al.}{2015}]{Atek2015}
{Atek} H.,  et~al., 2015, \mn@doi [\apj] {10.1088/0004-637X/814/1/69}, \href {https://ui.adsabs.harvard.edu/abs/2015ApJ...814...69A} {814, 69}

\bibitem[\protect\citeauthoryear{{Baldry} et~al.,}{{Baldry} et~al.}{2012}]{Baldry2012}
{Baldry} I.~K.,  et~al., 2012, \mn@doi [\mnras] {10.1111/j.1365-2966.2012.20340.x}, \href {https://ui.adsabs.harvard.edu/abs/2012MNRAS.421..621B} {421, 621}

\bibitem[\protect\citeauthoryear{{Battaglia} \& {Nipoti}}{{Battaglia} \& {Nipoti}}{2022}]{Battaglia2022}
{Battaglia} G.,  {Nipoti} C.,  2022, \mn@doi [Nature Astronomy] {10.1038/s41550-022-01638-7}, \href {https://ui.adsabs.harvard.edu/abs/2022NatAs...6..659B} {6, 659}

\bibitem[\protect\citeauthoryear{{Bennet}, {Sand}, {Crnojevi{\'c}}, {Spekkens}, {Karunakaran}, {Zaritsky}  \& {Mutlu-Pakdil}}{{Bennet} et~al.}{2020}]{Bennet2020}
{Bennet} P.,  {Sand} D.~J.,  {Crnojevi{\'c}} D.,  {Spekkens} K.,  {Karunakaran} A.,  {Zaritsky} D.,   {Mutlu-Pakdil} B.,  2020, \mn@doi [\apjl] {10.3847/2041-8213/ab80c5}, \href {https://ui.adsabs.harvard.edu/abs/2020ApJ...893L...9B} {893, L9}

\bibitem[\protect\citeauthoryear{{Bertin} \& {Arnouts}}{{Bertin} \& {Arnouts}}{1996}]{Bertin1996}
{Bertin} E.,  {Arnouts} S.,  1996, \mn@doi [\aaps] {10.1051/aas:1996164}, \href {https://ui.adsabs.harvard.edu/abs/1996A&AS..117..393B} {117, 393}

\bibitem[\protect\citeauthoryear{{Bichang'a}, {Kaviraj}, {Lazar}, {Jackson}, {Das}, {Smith}, {Watkins}  \& {Martin}}{{Bichang'a} et~al.}{2024}]{Bichanga2024}
{Bichang'a} B.,  {Kaviraj} S.,  {Lazar} I.,  {Jackson} R.~A.,  {Das} S.,  {Smith} D.~J.~B.,  {Watkins} A.~E.,   {Martin} G.,  2024, \mn@doi [\mnras] {10.1093/mnras/stae1441}, \href {https://ui.adsabs.harvard.edu/abs/2024MNRAS.532..613B} {532, 613}

\bibitem[\protect\citeauthoryear{{Bielby} et~al.,}{{Bielby} et~al.}{2012}]{Bielby2012}
{Bielby} R.,  et~al., 2012, \mn@doi [\aap] {10.1051/0004-6361/201118547}, \href {https://ui.adsabs.harvard.edu/abs/2012A&A...545A..23B} {545, A23}

\bibitem[\protect\citeauthoryear{{Bond}, {Kofman}  \& {Pogosyan}}{{Bond} et~al.}{1996}]{Bond1996}
{Bond} J.~R.,  {Kofman} L.,   {Pogosyan} D.,  1996, \mn@doi [\nat] {10.1038/380603a0}, \href {https://ui.adsabs.harvard.edu/abs/1996Natur.380..603B} {380, 603}

\bibitem[\protect\citeauthoryear{{Borch} et~al.,}{{Borch} et~al.}{2006}]{Borch2006}
{Borch} A.,  et~al., 2006, \mn@doi [\aap] {10.1051/0004-6361:20054376}, \href {https://ui.adsabs.harvard.edu/abs/2006A&A...453..869B} {453, 869}

\bibitem[\protect\citeauthoryear{{Boselli}, {Boissier}, {Cortese}  \& {Gavazzi}}{{Boselli} et~al.}{2008}]{Boselli2008}
{Boselli} A.,  {Boissier} S.,  {Cortese} L.,   {Gavazzi} G.,  2008, \mn@doi [\apj] {10.1086/525513}, \href {https://ui.adsabs.harvard.edu/#abs/2008ApJ...674..742B} {674}

\bibitem[\protect\citeauthoryear{{Boselli}, {Fossati}  \& {Sun}}{{Boselli} et~al.}{2022}]{Boselli2022}
{Boselli} A.,  {Fossati} M.,   {Sun} M.,  2022, \mn@doi [\aapr] {10.1007/s00159-022-00140-3}, \href {https://ui.adsabs.harvard.edu/abs/2022A&ARv..30....3B} {30, 3}

\bibitem[\protect\citeauthoryear{{Boylan-Kolchin}, {Bullock}  \& {Kaplinghat}}{{Boylan-Kolchin} et~al.}{2011}]{Boylan-Kolchin2011}
{Boylan-Kolchin} M.,  {Bullock} J.~S.,   {Kaplinghat} M.,  2011, \mn@doi [\mnras] {10.1111/j.1745-3933.2011.01074.x}, \href {https://ui.adsabs.harvard.edu/abs/2011MNRAS.415L..40B} {415, L40}

\bibitem[\protect\citeauthoryear{{Bullock} \& {Boylan-Kolchin}}{{Bullock} \& {Boylan-Kolchin}}{2017}]{Bullock2017}
{Bullock} J.~S.,  {Boylan-Kolchin} M.,  2017, \mn@doi [\araa] {10.1146/annurev-astro-091916-055313}, \href {https://ui.adsabs.harvard.edu/abs/2017ARA&A..55..343B} {55, 343}

\bibitem[\protect\citeauthoryear{{Choque-Challapa}, {Aguerri}, {Mancera Pi{\~n}a}, {Peletier}, {Venhola}  \& {Verheijen}}{{Choque-Challapa} et~al.}{2021}]{ChoqueChallapa2021}
{Choque-Challapa} N.,  {Aguerri} J. A.~L.,  {Mancera Pi{\~n}a} P.~E.,  {Peletier} R.,  {Venhola} A.,   {Verheijen} M.,  2021, \mn@doi [\mnras] {10.1093/mnras/stab2420}, \href {https://ui.adsabs.harvard.edu/abs/2021MNRAS.507.6045C} {507, 6045}

\bibitem[\protect\citeauthoryear{{Cuillandre} et~al.,}{{Cuillandre} et~al.}{2025}]{Cuillandre2025}
{Cuillandre} J.~C.,  et~al., 2025, \mn@doi [\aap] {10.1051/0004-6361/202450803}, \href {https://ui.adsabs.harvard.edu/abs/2025A&A...697A...6C} {697, A6}

\bibitem[\protect\citeauthoryear{{Dashyan}, {Silk}, {Mamon}, {Dubois}  \& {Hartwig}}{{Dashyan} et~al.}{2018}]{Dashyan2018}
{Dashyan} G.,  {Silk} J.,  {Mamon} G.~A.,  {Dubois} Y.,   {Hartwig} T.,  2018, \mn@doi [\mnras] {10.1093/mnras/stx2716}, \href {http://adsabs.harvard.edu/abs/2018MNRAS.473.5698D} {473, 5698}

\bibitem[\protect\citeauthoryear{{Davis} et~al.,}{{Davis} et~al.}{2022}]{Davis2022}
{Davis} F.,  et~al., 2022, \mn@doi [\mnras] {10.1093/mnras/stac068}, \href {https://ui.adsabs.harvard.edu/abs/2022MNRAS.511.4109D} {511, 4109}

\bibitem[\protect\citeauthoryear{{Desprez} et~al.,}{{Desprez} et~al.}{2023}]{Desprez2023}
{Desprez} G.,  et~al., 2023, \mn@doi [\aap] {10.1051/0004-6361/202243363}, \href {https://ui.adsabs.harvard.edu/abs/2023A&A...670A..82D} {670, A82}

\bibitem[\protect\citeauthoryear{{Dong}, {Lin}  \& {Murray}}{{Dong} et~al.}{2003}]{Dong2003}
{Dong} S.,  {Lin} D.~N.~C.,   {Murray} S.~D.,  2003, \mn@doi [\apj] {10.1086/378091}, \href {https://ui.adsabs.harvard.edu/abs/2003ApJ...596..930D} {596, 930}

\bibitem[\protect\citeauthoryear{{Drinkwater}, {Gregg}  \& {Colless}}{{Drinkwater} et~al.}{2001}]{Drinkwater2001}
{Drinkwater} M.~J.,  {Gregg} M.~D.,   {Colless} M.,  2001, \mn@doi [\apjl] {10.1086/319113}, \href {https://ui.adsabs.harvard.edu/abs/2001ApJ...548L.139D} {548, L139}

\bibitem[\protect\citeauthoryear{{Driver} et~al.,}{{Driver} et~al.}{2022}]{Driver2022}
{Driver} S.~P.,  et~al., 2022, \mn@doi [\mnras] {10.1093/mnras/stac472}, \href {https://ui.adsabs.harvard.edu/abs/2022MNRAS.513..439D} {513, 439}

\bibitem[\protect\citeauthoryear{{Dubois} et~al.,}{{Dubois} et~al.}{2014}]{Dubois2014}
{Dubois} Y.,  et~al., 2014, \mn@doi [\mnras] {10.1093/mnras/stu1227}, \href {http://adsabs.harvard.edu/abs/2014MNRAS.444.1453D} {444, 1453}

\bibitem[\protect\citeauthoryear{{Dubois} et~al.,}{{Dubois} et~al.}{2021}]{Dubois2021}
{Dubois} Y.,  et~al., 2021, \mn@doi [\aap] {10.1051/0004-6361/202039429}, \href {https://ui.adsabs.harvard.edu/abs/2021A&A...651A.109D} {651, A109}

\bibitem[\protect\citeauthoryear{{Emerick}, {Bryan}  \& {Mac Low}}{{Emerick} et~al.}{2019}]{Emerick2019}
{Emerick} A.,  {Bryan} G.~L.,   {Mac Low} M.-M.,  2019, \mn@doi [\mnras] {10.1093/mnras/sty2689}, \href {https://ui.adsabs.harvard.edu/abs/2019MNRAS.482.1304E} {482, 1304}

\bibitem[\protect\citeauthoryear{{Etherington} et~al.,}{{Etherington} et~al.}{2017}]{Etherington2017}
{Etherington} J.,  et~al., 2017, \mn@doi [\mnras] {10.1093/mnras/stw3069}, \href {https://ui.adsabs.harvard.edu/abs/2017MNRAS.466..228E} {466, 228}

\bibitem[\protect\citeauthoryear{{Fattahi}, {Navarro}, {Frenk}, {Oman}, {Sawala}  \& {Schaller}}{{Fattahi} et~al.}{2018}]{Fattahi2018}
{Fattahi} A.,  {Navarro} J.~F.,  {Frenk} C.~S.,  {Oman} K.~A.,  {Sawala} T.,   {Schaller} M.,  2018, \mn@doi [\mnras] {10.1093/mnras/sty408}, \href {https://ui.adsabs.harvard.edu/abs/2018MNRAS.476.3816F} {476, 3816}

\bibitem[\protect\citeauthoryear{{Feldmann} et~al.,}{{Feldmann} et~al.}{2023}]{Feldmann2023}
{Feldmann} R.,  et~al., 2023, \mn@doi [\mnras] {10.1093/mnras/stad1205}, \href {https://ui.adsabs.harvard.edu/abs/2023MNRAS.522.3831F} {522, 3831}

\bibitem[\protect\citeauthoryear{{Finoguenov} et~al.,}{{Finoguenov} et~al.}{2007}]{Finoguenov2007}
{Finoguenov} A.,  et~al., 2007, \mn@doi [\apjs] {10.1086/516577}, \href {https://ui.adsabs.harvard.edu/abs/2007ApJS..172..182F} {172, 182}

\bibitem[\protect\citeauthoryear{{Fontana} et~al.,}{{Fontana} et~al.}{2004}]{Fontana2004}
{Fontana} A.,  et~al., 2004, \mn@doi [\aap] {10.1051/0004-6361:20035626}, \href {https://ui.adsabs.harvard.edu/abs/2004A&A...424...23F} {424, 23}

\bibitem[\protect\citeauthoryear{{Forbes}, {Krumholz}, {Goldbaum}  \& {Dekel}}{{Forbes} et~al.}{2016}]{Forbes2016}
{Forbes} J.~C.,  {Krumholz} M.~R.,  {Goldbaum} N.~J.,   {Dekel} A.,  2016, \mn@doi [\nat] {10.1038/nature18292}, \href {https://ui.adsabs.harvard.edu/abs/2016Natur.535..523F} {535, 523}

\bibitem[\protect\citeauthoryear{{Fouqu{\'e}}, {Solanes}, {Sanchis}  \& {Balkowski}}{{Fouqu{\'e}} et~al.}{2001}]{Fouque2001}
{Fouqu{\'e}} P.,  {Solanes} J.~M.,  {Sanchis} T.,   {Balkowski} C.,  2001, \mn@doi [\aap] {10.1051/0004-6361:20010833}, \href {https://ui.adsabs.harvard.edu/abs/2001A&A...375..770F} {375, 770}

\bibitem[\protect\citeauthoryear{{Gavazzi}, {Adami}, {Durret}, {Cuillandre}, {Ilbert}, {Mazure}, {Pell{\'o}}  \& {Ulmer}}{{Gavazzi} et~al.}{2009}]{Gavazzi2009}
{Gavazzi} R.,  {Adami} C.,  {Durret} F.,  {Cuillandre} J.~C.,  {Ilbert} O.,  {Mazure} A.,  {Pell{\'o}} R.,   {Ulmer} M.~P.,  2009, \mn@doi [\aap] {10.1051/0004-6361/200911841}, \href {https://ui.adsabs.harvard.edu/abs/2009A&A...498L..33G} {498, L33}

\bibitem[\protect\citeauthoryear{{Geen}, {Slyz}  \& {Devriendt}}{{Geen} et~al.}{2013}]{Geen2013}
{Geen} S.,  {Slyz} A.,   {Devriendt} J.,  2013, \mn@doi [\mnras] {10.1093/mnras/sts364}, \href {https://ui.adsabs.harvard.edu/abs/2013MNRAS.429..633G} {429, 633}

\bibitem[\protect\citeauthoryear{{Geha} et~al.,}{{Geha} et~al.}{2024}]{Geha2024}
{Geha} M.,  et~al., 2024, \mn@doi [\apj] {10.3847/1538-4357/ad61e7}, \href {https://ui.adsabs.harvard.edu/abs/2024ApJ...976..118G} {976, 118}

\bibitem[\protect\citeauthoryear{{George} et~al.,}{{George} et~al.}{2011}]{George2011}
{George} M.~R.,  et~al., 2011, \mn@doi [\apj] {10.1088/0004-637X/742/2/125}, \href {https://ui.adsabs.harvard.edu/abs/2011ApJ...742..125G} {742, 125}

\bibitem[\protect\citeauthoryear{{Gozaliasl} et~al.,}{{Gozaliasl} et~al.}{2014}]{Gozaliasl2014}
{Gozaliasl} G.,  et~al., 2014, \mn@doi [\aap] {10.1051/0004-6361/201322459}, \href {https://ui.adsabs.harvard.edu/abs/2014A&A...566A.140G} {566, A140}

\bibitem[\protect\citeauthoryear{{Gozaliasl} et~al.,}{{Gozaliasl} et~al.}{2019}]{Gozaliasl2019}
{Gozaliasl} G.,  et~al., 2019, \mn@doi [\mnras] {10.1093/mnras/sty3203}, \href {https://ui.adsabs.harvard.edu/abs/2019MNRAS.483.3545G} {483, 3545}

\bibitem[\protect\citeauthoryear{{Hasan} et~al.,}{{Hasan} et~al.}{2023}]{Hasan2023}
{Hasan} F.,  et~al., 2023, \mn@doi [\apj] {10.3847/1538-4357/acd11c}, \href {https://ui.adsabs.harvard.edu/abs/2023ApJ...950..114H} {950, 114}

\bibitem[\protect\citeauthoryear{{Hazenfratz}, {Barai}, {Lanfranchi}  \& {Caproni}}{{Hazenfratz} et~al.}{2024}]{Hazenfratz2024}
{Hazenfratz} R.,  {Barai} P.,  {Lanfranchi} G.~A.,   {Caproni} A.,  2024, \mn@doi [\apj] {10.3847/1538-4357/ad4700}, \href {https://ui.adsabs.harvard.edu/abs/2024ApJ...969...65H} {969, 65}

\bibitem[\protect\citeauthoryear{{Hopkins}}{{Hopkins}}{2015}]{Hopkins2015}
{Hopkins} P.~F.,  2015, \mn@doi [\mnras] {10.1093/mnras/stv195}, \href {https://ui.adsabs.harvard.edu/abs/2015MNRAS.450...53H} {450, 53}

\bibitem[\protect\citeauthoryear{{Hopkins} et~al.,}{{Hopkins} et~al.}{2018}]{Hopkins2018}
{Hopkins} P.~F.,  et~al., 2018, \mn@doi [\mnras] {10.1093/mnras/sty1690}, \href {https://ui.adsabs.harvard.edu/abs/2018MNRAS.480..800H} {480, 800}

\bibitem[\protect\citeauthoryear{{Hsieh}, {Wang}, {Hsieh}, {Lin}, {Yan}, {Lim}  \& {Ho}}{{Hsieh} et~al.}{2012}]{Hsieh2012}
{Hsieh} B.-C.,  {Wang} W.-H.,  {Hsieh} C.-C.,  {Lin} L.,  {Yan} H.,  {Lim} J.,   {Ho} P. T.~P.,  2012, \mn@doi [\apjs] {10.1088/0067-0049/203/2/23}, \href {https://ui.adsabs.harvard.edu/abs/2012ApJS..203...23H} {203, 23}

\bibitem[\protect\citeauthoryear{{Hu}, {Barkana}  \& {Gruzinov}}{{Hu} et~al.}{2000}]{Hu2000}
{Hu} W.,  {Barkana} R.,   {Gruzinov} A.,  2000, \mn@doi [\prl] {10.1103/PhysRevLett.85.1158}, \href {https://ui.adsabs.harvard.edu/abs/2000PhRvL..85.1158H} {85, 1158}

\bibitem[\protect\citeauthoryear{{Hubble}}{{Hubble}}{1936}]{Hubble1936}
{Hubble} E.~P.,  1936, {Realm of the Nebulae}

\bibitem[\protect\citeauthoryear{{Ilbert} et~al.,}{{Ilbert} et~al.}{2006}]{Ilbert2006}
{Ilbert} O.,  et~al., 2006, \mn@doi [\aap] {10.1051/0004-6361:20065138}, \href {https://ui.adsabs.harvard.edu/abs/2006A&A...457..841I} {457, 841}

\bibitem[\protect\citeauthoryear{{Ivezi{\'c}} et~al.,}{{Ivezi{\'c}} et~al.}{2019}]{Ivezic2019}
{Ivezi{\'c}} {\v{Z}}.,  et~al., 2019, \mn@doi [\apj] {10.3847/1538-4357/ab042c}, \href {https://ui.adsabs.harvard.edu/abs/2019ApJ...873..111I} {873, 111}

\bibitem[\protect\citeauthoryear{{Jackson} et~al.,}{{Jackson} et~al.}{2021a}]{Jackson2021a}
{Jackson} R.~A.,  et~al., 2021a, \mn@doi [\mnras] {10.1093/mnras/stab093}, \href {https://ui.adsabs.harvard.edu/abs/2021MNRAS.502.1785J} {502, 1785}

\bibitem[\protect\citeauthoryear{{Jackson} et~al.,}{{Jackson} et~al.}{2021b}]{Jackson2021b}
{Jackson} R.~A.,  et~al., 2021b, \mn@doi [\mnras] {10.1093/mnras/stab077}, \href {https://ui.adsabs.harvard.edu/abs/2021MNRAS.502.4262J} {502, 4262}

\bibitem[\protect\citeauthoryear{{Jeon} et~al.,}{{Jeon} et~al.}{2025}]{Jeon2025}
{Jeon} S.,  et~al., 2025, \mn@doi [arXiv e-prints] {10.48550/arXiv.2506.09152}, \href {https://ui.adsabs.harvard.edu/abs/2025arXiv250609152J} {p. arXiv:2506.09152}

\bibitem[\protect\citeauthoryear{{Jespersen}, {Steinhardt}, {Somerville}  \& {Lovell}}{{Jespersen} et~al.}{2025}]{Jespersen2025}
{Jespersen} C.~K.,  {Steinhardt} C.~L.,  {Somerville} R.~S.,   {Lovell} C.~C.,  2025, \mn@doi [\apj] {10.3847/1538-4357/adb422}, \href {https://ui.adsabs.harvard.edu/abs/2025ApJ...982...23J} {982, 23}

\bibitem[\protect\citeauthoryear{{Kaiser}}{{Kaiser}}{1984}]{Kaiser1984}
{Kaiser} N.,  1984, \mn@doi [\apjl] {10.1086/184341}, \href {https://ui.adsabs.harvard.edu/abs/1984ApJ...284L...9K} {284, L9}

\bibitem[\protect\citeauthoryear{{Kaviraj} et~al.,}{{Kaviraj} et~al.}{2017}]{Kaviraj2017}
{Kaviraj} S.,  et~al., 2017, \mn@doi [\mnras] {10.1093/mnras/stx126}, \href {http://adsabs.harvard.edu/abs/2017MNRAS.467.4739K} {467, 4739}

\bibitem[\protect\citeauthoryear{{Kaviraj}, {Martin}  \& {Silk}}{{Kaviraj} et~al.}{2019}]{Kaviraj2019}
{Kaviraj} S.,  {Martin} G.,   {Silk} J.,  2019, \mn@doi [\mnras] {10.1093/mnrasl/slz102}, \href {https://ui.adsabs.harvard.edu/abs/2019MNRAS.489L..12K} {489, L12}

\bibitem[\protect\citeauthoryear{{Kaviraj}, {Lazar}, {Watkins}, {Laigle}, {Martin}  \& {Jackson}}{{Kaviraj} et~al.}{2025}]{Kaviraj2025}
{Kaviraj} S.,  {Lazar} I.,  {Watkins} A.~E.,  {Laigle} C.,  {Martin} G.,   {Jackson} R.~A.,  2025, \mn@doi [\mnras] {10.1093/mnras/staf233}, \href {https://ui.adsabs.harvard.edu/abs/2025MNRAS.538..153K} {538, 153}

\bibitem[\protect\citeauthoryear{{Kennicutt}}{{Kennicutt}}{1998}]{Kennicutt1998}
{Kennicutt} Jr. R.~C.,  1998, \mn@doi [\apj] {10.1086/305588}, \href {http://adsabs.harvard.edu/abs/1998ApJ...498..541K} {498, 541}

\bibitem[\protect\citeauthoryear{{Kimm} \& {Cen}}{{Kimm} \& {Cen}}{2014}]{Kimm2014}
{Kimm} T.,  {Cen} R.,  2014, \mn@doi [\apj] {10.1088/0004-637X/788/2/121}, \href {https://ui.adsabs.harvard.edu/abs/2014ApJ...788..121K} {788, 121}

\bibitem[\protect\citeauthoryear{{Klypin}, {Kravtsov}, {Valenzuela}  \& {Prada}}{{Klypin} et~al.}{1999}]{Klypin1999}
{Klypin} A.,  {Kravtsov} A.~V.,  {Valenzuela} O.,   {Prada} F.,  1999, \mn@doi [\apj] {10.1086/307643}, \href {https://ui.adsabs.harvard.edu/abs/1999ApJ...522...82K} {522, 82}

\bibitem[\protect\citeauthoryear{{Koch}, {Burkert}, {Rich}, {Collins}, {Black}, {Hilker}  \& {Benson}}{{Koch} et~al.}{2012}]{Koch2012}
{Koch} A.,  {Burkert} A.,  {Rich} R.~M.,  {Collins} M. L.~M.,  {Black} C.~S.,  {Hilker} M.,   {Benson} A.~J.,  2012, \mn@doi [\apjl] {10.1088/2041-8205/755/1/L13}, \href {https://ui.adsabs.harvard.edu/abs/2012ApJ...755L..13K} {755, L13}

\bibitem[\protect\citeauthoryear{{Kormendy}}{{Kormendy}}{1985}]{Kormendy1985}
{Kormendy} J.,  1985, \mn@doi [\apj] {10.1086/163350}, \href {https://ui.adsabs.harvard.edu/abs/1985ApJ...295...73K} {295, 73}

\bibitem[\protect\citeauthoryear{{Koudmani}, {Henden}  \& {Sijacki}}{{Koudmani} et~al.}{2021}]{Koudmani2021}
{Koudmani} S.,  {Henden} N.~A.,   {Sijacki} D.,  2021, \mn@doi [\mnras] {10.1093/mnras/stab677}, \href {https://ui.adsabs.harvard.edu/abs/2021MNRAS.tmp..682K} {}

\bibitem[\protect\citeauthoryear{{Koudmani}, {Sijacki}  \& {Smith}}{{Koudmani} et~al.}{2022}]{Koudmani2022}
{Koudmani} S.,  {Sijacki} D.,   {Smith} M.~C.,  2022, \mn@doi [\mnras] {10.1093/mnras/stac2252}, \href {https://ui.adsabs.harvard.edu/abs/2022MNRAS.516.2112K} {516, 2112}

\bibitem[\protect\citeauthoryear{{Kraljic} et~al.,}{{Kraljic} et~al.}{2018}]{Kraljic2018}
{Kraljic} K.,  et~al., 2018, \mn@doi [\mnras] {10.1093/mnras/stx2638}, \href {https://ui.adsabs.harvard.edu/abs/2018MNRAS.474..547K} {474, 547}

\bibitem[\protect\citeauthoryear{{Kraljic} et~al.,}{{Kraljic} et~al.}{2019}]{Kraljic2019}
{Kraljic} K.,  et~al., 2019, \mn@doi [\mnras] {10.1093/mnras/sty3216}, \href {https://ui.adsabs.harvard.edu/abs/2019MNRAS.483.3227K} {483, 3227}

\bibitem[\protect\citeauthoryear{{Laigle} et~al.,}{{Laigle} et~al.}{2018}]{Laigle2018}
{Laigle} C.,  et~al., 2018, \mn@doi [\mnras] {10.1093/mnras/stx3055}, \href {https://ui.adsabs.harvard.edu/abs/2018MNRAS.474.5437L} {474, 5437}

\bibitem[\protect\citeauthoryear{{Laigle} et~al.,}{{Laigle} et~al.}{2019}]{Laigle2019}
{Laigle} C.,  et~al., 2019, \mn@doi [\mnras] {10.1093/mnras/stz1054}, \href {https://ui.adsabs.harvard.edu/abs/2019MNRAS.486.5104L} {486, 5104}

\bibitem[\protect\citeauthoryear{{Laureijs} et~al.,}{{Laureijs} et~al.}{2011}]{Laureijs2011}
{Laureijs} R.,  et~al., 2011, preprint, \href {http://adsabs.harvard.edu/abs/2011arXiv1110.3193L} {} (\mn@eprint {arXiv} {1110.3193})

\bibitem[\protect\citeauthoryear{{Lazar}, {Kaviraj}, {Martin}, {Laigle}, {Watkins}  \& {Jackson}}{{Lazar} et~al.}{2023}]{Lazar2023}
{Lazar} I.,  {Kaviraj} S.,  {Martin} G.,  {Laigle} C.,  {Watkins} A.,   {Jackson} R.~A.,  2023, \mn@doi [\mnras] {10.1093/mnras/stad224}, \href {https://ui.adsabs.harvard.edu/abs/2023MNRAS.520.2109L} {520, 2109}

\bibitem[\protect\citeauthoryear{{Lazar}, {Kaviraj}, {Watkins}, {Martin}, {Bichang'a}  \& {Jackson}}{{Lazar} et~al.}{2024a}]{Lazar2024a}
{Lazar} I.,  {Kaviraj} S.,  {Watkins} A.~E.,  {Martin} G.,  {Bichang'a} B.,   {Jackson} R.~A.,  2024a, \mn@doi [\mnras] {10.1093/mnras/stae510}, \href {https://ui.adsabs.harvard.edu/abs/2024MNRAS.529..499L} {529, 499}

\bibitem[\protect\citeauthoryear{{Lazar}, {Kaviraj}, {Watkins}, {Martin}, {Bichang'a}  \& {Jackson}}{{Lazar} et~al.}{2024b}]{Lazar2024b}
{Lazar} I.,  {Kaviraj} S.,  {Watkins} A.~E.,  {Martin} G.,  {Bichang'a} B.,   {Jackson} R.~A.,  2024b, \mn@doi [\mnras] {10.1093/mnras/stae1956}, \href {https://ui.adsabs.harvard.edu/abs/2024MNRAS.533.3771L} {533, 3771}

\bibitem[\protect\citeauthoryear{{Leauthaud} et~al.,}{{Leauthaud} et~al.}{2007}]{Leauthaud2007}
{Leauthaud} A.,  et~al., 2007, \mn@doi [\apjs] {10.1086/516598}, \href {https://ui.adsabs.harvard.edu/abs/2007ApJS..172..219L} {172, 219}

\bibitem[\protect\citeauthoryear{{Lovell} et~al.,}{{Lovell} et~al.}{2012}]{Lovell2012}
{Lovell} M.~R.,  et~al., 2012, \mn@doi [\mnras] {10.1111/j.1365-2966.2011.20200.x}, \href {https://ui.adsabs.harvard.edu/abs/2012MNRAS.420.2318L} {420, 2318}

\bibitem[\protect\citeauthoryear{{Malavasi}, {Pozzetti}, {Cucciati}, {Bardelli}, {Ilbert}  \& {Cimatti}}{{Malavasi} et~al.}{2017}]{Malavasi2017}
{Malavasi} N.,  {Pozzetti} L.,  {Cucciati} O.,  {Bardelli} S.,  {Ilbert} O.,   {Cimatti} A.,  2017, \mn@doi [\mnras] {10.1093/mnras/stx1323}, \href {https://ui.adsabs.harvard.edu/abs/2017MNRAS.470.1274M} {470, 1274}

\bibitem[\protect\citeauthoryear{{Marsh}}{{Marsh}}{2016}]{Marsh2016}
{Marsh} D. J.~E.,  2016, \mn@doi [arXiv e-prints] {10.48550/arXiv.1605.05973}, \href {https://ui.adsabs.harvard.edu/abs/2016arXiv160505973M} {p. arXiv:1605.05973}

\bibitem[\protect\citeauthoryear{{Martel}, {Barai}  \& {Brito}}{{Martel} et~al.}{2012}]{Martel2012}
{Martel} H.,  {Barai} P.,   {Brito} W.,  2012, \mn@doi [\apj] {10.1088/0004-637X/757/1/48}, \href {https://ui.adsabs.harvard.edu/abs/2012ApJ...757...48M} {757, 48}

\bibitem[\protect\citeauthoryear{{Martin}, {Pearce}, {Hatch}, {Contreras-Santos}, {Knebe}  \& {Cui}}{{Martin} et~al.}{2024}]{Martin2024}
{Martin} G.,  {Pearce} F.~R.,  {Hatch} N.~A.,  {Contreras-Santos} A.,  {Knebe} A.,   {Cui} W.,  2024, \mn@doi [\mnras] {10.1093/mnras/stae2488}, \href {https://ui.adsabs.harvard.edu/abs/2024MNRAS.535.2375M} {535, 2375}

\bibitem[\protect\citeauthoryear{{Martin} et~al.,}{{Martin} et~al.}{2025}]{Martin2025}
{Martin} G.,  et~al., 2025, \mn@doi [arXiv e-prints] {10.48550/arXiv.2505.04509}, \href {https://ui.adsabs.harvard.edu/abs/2025arXiv250504509M} {p. arXiv:2505.04509}

\bibitem[\protect\citeauthoryear{{Mayer}, {Mastropietro}, {Wadsley}, {Stadel}  \& {Moore}}{{Mayer} et~al.}{2006}]{Mayer2006}
{Mayer} L.,  {Mastropietro} C.,  {Wadsley} J.,  {Stadel} J.,   {Moore} B.,  2006, \mn@doi [\mnras] {10.1111/j.1365-2966.2006.10403.x}, \href {https://ui.adsabs.harvard.edu/abs/2006MNRAS.369.1021M} {369, 1021}

\bibitem[\protect\citeauthoryear{{McCracken} et~al.,}{{McCracken} et~al.}{2012}]{McCracken2012}
{McCracken} H.~J.,  et~al., 2012, \mn@doi [\aap] {10.1051/0004-6361/201219507}, \href {https://ui.adsabs.harvard.edu/abs/2012A&A...544A.156M} {544, A156}

\bibitem[\protect\citeauthoryear{{Mezcua} \& {Dom{\'\i}nguez S{\'a}nchez}}{{Mezcua} \& {Dom{\'\i}nguez S{\'a}nchez}}{2024}]{Mezcua2024}
{Mezcua} M.,  {Dom{\'\i}nguez S{\'a}nchez} H.,  2024, \mn@doi [\mnras] {10.1093/mnras/stae292}, \href {https://ui.adsabs.harvard.edu/abs/2024MNRAS.528.5252M} {528, 5252}

\bibitem[\protect\citeauthoryear{{Moore}, {Ghigna}, {Governato}, {Lake}, {Quinn}, {Stadel}  \& {Tozzi}}{{Moore} et~al.}{1999}]{Moore1999}
{Moore} B.,  {Ghigna} S.,  {Governato} F.,  {Lake} G.,  {Quinn} T.,  {Stadel} J.,   {Tozzi} P.,  1999, \mn@doi [\apjl] {10.1086/312287}, \href {https://ui.adsabs.harvard.edu/abs/1999ApJ...524L..19M} {524, L19}

\bibitem[\protect\citeauthoryear{{Mori} \& {Burkert}}{{Mori} \& {Burkert}}{2000}]{Mori2000}
{Mori} M.,  {Burkert} A.,  2000, \mn@doi [\apj] {10.1086/309140}, \href {https://ui.adsabs.harvard.edu/abs/2000ApJ...538..559M} {538, 559}

\bibitem[\protect\citeauthoryear{{Mortlock} et~al.,}{{Mortlock} et~al.}{2015}]{Mortlock2015}
{Mortlock} A.,  et~al., 2015, \mn@doi [\mnras] {10.1093/mnras/stu2403}, \href {https://ui.adsabs.harvard.edu/abs/2015MNRAS.447....2M} {447, 2}

\bibitem[\protect\citeauthoryear{{Moster}, {Somerville}, {Newman}  \& {Rix}}{{Moster} et~al.}{2011}]{Moster2010}
{Moster} B.~P.,  {Somerville} R.~S.,  {Newman} J.~A.,   {Rix} H.-W.,  2011, \mn@doi [\apj] {10.1088/0004-637X/731/2/113}, \href {https://ui.adsabs.harvard.edu/abs/2011ApJ...731..113M} {731, 113}

\bibitem[\protect\citeauthoryear{{Moustakas} et~al.,}{{Moustakas} et~al.}{2013}]{Moustakas2013}
{Moustakas} J.,  et~al., 2013, \mn@doi [\apj] {10.1088/0004-637X/767/1/50}, \href {https://ui.adsabs.harvard.edu/abs/2013ApJ...767...50M} {767, 50}

\bibitem[\protect\citeauthoryear{{Musso}, {Cadiou}, {Pichon}, {Codis}, {Kraljic}  \& {Dubois}}{{Musso} et~al.}{2018}]{Musso2018}
{Musso} M.,  {Cadiou} C.,  {Pichon} C.,  {Codis} S.,  {Kraljic} K.,   {Dubois} Y.,  2018, \mn@doi [\mnras] {10.1093/mnras/sty191}, \href {https://ui.adsabs.harvard.edu/abs/2018MNRAS.476.4877M} {476, 4877}

\bibitem[\protect\citeauthoryear{{Nashimoto}, {Tanaka}, {Chiba}, {Hayashi}, {Komiyama}  \& {Okamoto}}{{Nashimoto} et~al.}{2022}]{Nashimoto2022}
{Nashimoto} M.,  {Tanaka} M.,  {Chiba} M.,  {Hayashi} K.,  {Komiyama} Y.,   {Okamoto} T.,  2022, \mn@doi [\apj] {10.3847/1538-4357/ac83a4}, \href {https://ui.adsabs.harvard.edu/abs/2022ApJ...936...38N} {936, 38}

\bibitem[\protect\citeauthoryear{{Nelson} et~al.,}{{Nelson} et~al.}{2019}]{Nelson2019}
{Nelson} D.,  et~al., 2019, \mn@doi [\mnras] {10.1093/mnras/stz2306}, \href {https://ui.adsabs.harvard.edu/abs/2019MNRAS.490.3234N} {490, 3234}

\bibitem[\protect\citeauthoryear{{Obreschkow}, {Murray}, {Robotham}  \& {Westmeier}}{{Obreschkow} et~al.}{2018}]{Obreschkow2018}
{Obreschkow} D.,  {Murray} S.~G.,  {Robotham} A.~S.~G.,   {Westmeier} T.,  2018, \mn@doi [\mnras] {10.1093/mnras/stx3155}, \href {https://ui.adsabs.harvard.edu/abs/2018MNRAS.474.5500O} {474, 5500}

\bibitem[\protect\citeauthoryear{{Picouet} et~al.,}{{Picouet} et~al.}{2023}]{Picouet2023}
{Picouet} V.,  et~al., 2023, \mn@doi [\aap] {10.1051/0004-6361/202245756}, \href {https://ui.adsabs.harvard.edu/abs/2023A&A...675A.164P} {675, A164}

\bibitem[\protect\citeauthoryear{{Pillepich} et~al.,}{{Pillepich} et~al.}{2019}]{Pillepich2019}
{Pillepich} A.,  et~al., 2019, \mn@doi [\mnras] {10.1093/mnras/stz2338}, \href {https://ui.adsabs.harvard.edu/abs/2019MNRAS.490.3196P} {490, 3196}

\bibitem[\protect\citeauthoryear{{Polisensky} \& {Ricotti}}{{Polisensky} \& {Ricotti}}{2011}]{Polisensky2011}
{Polisensky} E.,  {Ricotti} M.,  2011, \mn@doi [\prd] {10.1103/PhysRevD.83.043506}, \href {https://ui.adsabs.harvard.edu/abs/2011PhRvD..83d3506P} {83, 043506}

\bibitem[\protect\citeauthoryear{{Poulain} et~al.,}{{Poulain} et~al.}{2021}]{Poulain2021}
{Poulain} M.,  et~al., 2021, \mn@doi [\mnras] {10.1093/mnras/stab2092}, \href {https://ui.adsabs.harvard.edu/abs/2021MNRAS.506.5494P} {506, 5494}

\bibitem[\protect\citeauthoryear{{Pozzetti} et~al.,}{{Pozzetti} et~al.}{2007}]{Pozzetti2007}
{Pozzetti} L.,  et~al., 2007, \mn@doi [\aap] {10.1051/0004-6361:20077609}, \href {https://ui.adsabs.harvard.edu/abs/2007A&A...474..443P} {474, 443}

\bibitem[\protect\citeauthoryear{{Reaves}}{{Reaves}}{1956}]{reaves56}
{Reaves} G.,  1956, \mn@doi [\aj] {10.1086/107292}, \href {https://ui.adsabs.harvard.edu/abs/1956AJ.....61...69R} {61, 69}

\bibitem[\protect\citeauthoryear{{Salvadori}, {Sk{\'u}lad{\'o}ttir}  \& {Tolstoy}}{{Salvadori} et~al.}{2015}]{Salvadori2015}
{Salvadori} S.,  {Sk{\'u}lad{\'o}ttir} {\'A}.,   {Tolstoy} E.,  2015, \mn@doi [\mnras] {10.1093/mnras/stv1969}, \href {https://ui.adsabs.harvard.edu/abs/2015MNRAS.454.1320S} {454, 1320}

\bibitem[\protect\citeauthoryear{{Santos-Santos}, {Sales}, {Fattahi}  \& {Navarro}}{{Santos-Santos} et~al.}{2022}]{Santos-Santos2022}
{Santos-Santos} I. M.~E.,  {Sales} L.~V.,  {Fattahi} A.,   {Navarro} J.~F.,  2022, \mn@doi [\mnras] {10.1093/mnras/stac2057}, \href {https://ui.adsabs.harvard.edu/abs/2022MNRAS.515.3685S} {515, 3685}

\bibitem[\protect\citeauthoryear{{Sawala} et~al.,}{{Sawala} et~al.}{2016}]{Sawala2016}
{Sawala} T.,  et~al., 2016, \mn@doi [\mnras] {10.1093/mnras/stw145}, \href {https://ui.adsabs.harvard.edu/abs/2016MNRAS.457.1931S} {457, 1931}

\bibitem[\protect\citeauthoryear{{Sawicki} et~al.,}{{Sawicki} et~al.}{2019}]{Sawicki2019}
{Sawicki} M.,  et~al., 2019, \mn@doi [\mnras] {10.1093/mnras/stz2522}, \href {https://ui.adsabs.harvard.edu/abs/2019MNRAS.489.5202S} {489, 5202}

\bibitem[\protect\citeauthoryear{{Schaap} \& {van de Weygaert}}{{Schaap} \& {van de Weygaert}}{2000}]{Schaap2000}
{Schaap} W.~E.,  {van de Weygaert} R.,  2000, \mn@doi [\aap] {10.48550/arXiv.astro-ph/0011007}, \href {https://ui.adsabs.harvard.edu/abs/2000A&A...363L..29S} {363, L29}

\bibitem[\protect\citeauthoryear{{Schechter}}{{Schechter}}{1976}]{Schechter1976}
{Schechter} P.,  1976, \mn@doi [\apj] {10.1086/154079}, \href {https://ui.adsabs.harvard.edu/#abs/1976ApJ...203..297S} {203, 297}

\bibitem[\protect\citeauthoryear{{Scoville} et~al.,}{{Scoville} et~al.}{2007}]{Scoville2007}
{Scoville} N.,  et~al., 2007, \mn@doi [\apjs] {10.1086/516585}, \href {https://ui.adsabs.harvard.edu/abs/2007ApJS..172....1S} {172, 1}

\bibitem[\protect\citeauthoryear{{Shapley}}{{Shapley}}{1938}]{shapley38}
{Shapley} H.,  1938, Harvard College Observatory Bulletin, \href {https://ui.adsabs.harvard.edu/abs/1938BHarO.908....1S} {908, 1}

\bibitem[\protect\citeauthoryear{{Silk}}{{Silk}}{2017}]{Silk2017}
{Silk} J.,  2017, \mn@doi [\apjl] {10.3847/2041-8213/aa67da}, \href {http://adsabs.harvard.edu/abs/2017ApJ...839L..13S} {839, L13}

\bibitem[\protect\citeauthoryear{{Smercina}, {Bell}, {Price}, {D'Souza}, {Slater}, {Bailin}, {Monachesi}  \& {Nidever}}{{Smercina} et~al.}{2018}]{Smercina2018}
{Smercina} A.,  {Bell} E.~F.,  {Price} P.~A.,  {D'Souza} R.,  {Slater} C.~T.,  {Bailin} J.,  {Monachesi} A.,   {Nidever} D.,  2018, \mn@doi [\apj] {10.3847/1538-4357/aad2d6}, \href {https://ui.adsabs.harvard.edu/abs/2018ApJ...863..152S} {863, 152}

\bibitem[\protect\citeauthoryear{{Song} et~al.,}{{Song} et~al.}{2021}]{Song2021}
{Song} H.,  et~al., 2021, \mn@doi [\mnras] {10.1093/mnras/staa3981}, \href {https://ui.adsabs.harvard.edu/abs/2021MNRAS.501.4635S} {501, 4635}

\bibitem[\protect\citeauthoryear{{Sousbie}}{{Sousbie}}{2011}]{Sousbie2011}
{Sousbie} T.,  2011, \mn@doi [\mnras] {10.1111/j.1365-2966.2011.18394.x}, \href {https://ui.adsabs.harvard.edu/abs/2011MNRAS.414..350S} {414, 350}

\bibitem[\protect\citeauthoryear{{Springel}}{{Springel}}{2010}]{Springel2010}
{Springel} V.,  2010, \mn@doi [\mnras] {10.1111/j.1365-2966.2009.15715.x}, \href {https://ui.adsabs.harvard.edu/abs/2010MNRAS.401..791S} {401, 791}

\bibitem[\protect\citeauthoryear{{Springel} \& {Hernquist}}{{Springel} \& {Hernquist}}{2003}]{Springel2003}
{Springel} V.,  {Hernquist} L.,  2003, \mn@doi [\mnras] {10.1046/j.1365-8711.2003.06206.x}, \href {https://ui.adsabs.harvard.edu/abs/2003MNRAS.339..289S} {339, 289}

\bibitem[\protect\citeauthoryear{{Steinhardt} et~al.,}{{Steinhardt} et~al.}{2014}]{Steinhardt2014}
{Steinhardt} C.~L.,  et~al., 2014, \mn@doi [\apjl] {10.1088/2041-8205/791/2/L25}, \href {https://ui.adsabs.harvard.edu/abs/2014ApJ...791L..25S} {791, L25}

\bibitem[\protect\citeauthoryear{{Steyrleithner}, {Hensler}  \& {Boselli}}{{Steyrleithner} et~al.}{2020}]{Steyrleithner2020}
{Steyrleithner} P.,  {Hensler} G.,   {Boselli} A.,  2020, \mn@doi [\mnras] {10.1093/mnras/staa775}, \href {https://ui.adsabs.harvard.edu/abs/2020MNRAS.494.1114S} {494, 1114}

\bibitem[\protect\citeauthoryear{{Tanaka}, {Chiba}, {Hayashi}, {Komiyama}, {Okamoto}, {Cooper}, {Okamoto}  \& {Spitler}}{{Tanaka} et~al.}{2018}]{Tanaka2018}
{Tanaka} M.,  {Chiba} M.,  {Hayashi} K.,  {Komiyama} Y.,  {Okamoto} T.,  {Cooper} A.~P.,  {Okamoto} S.,   {Spitler} L.,  2018, \mn@doi [\apj] {10.3847/1538-4357/aad9fe}, \href {https://ui.adsabs.harvard.edu/abs/2018ApJ...865..125T} {865, 125}

\bibitem[\protect\citeauthoryear{{Taniguchi} et~al.,}{{Taniguchi} et~al.}{2007}]{Taniguchi2007}
{Taniguchi} Y.,  et~al., 2007, \mn@doi [\apjs] {10.1086/516596}, \href {https://ui.adsabs.harvard.edu/abs/2007ApJS..172....9T} {172, 9}

\bibitem[\protect\citeauthoryear{{Taniguchi} et~al.,}{{Taniguchi} et~al.}{2015}]{Taniguchi2015}
{Taniguchi} Y.,  et~al., 2015, \mn@doi [\pasj] {10.1093/pasj/psv106}, \href {https://ui.adsabs.harvard.edu/abs/2015PASJ...67..104T} {67, 104}

\bibitem[\protect\citeauthoryear{{Teyssier}}{{Teyssier}}{2002}]{Teyssier2002}
{Teyssier} R.,  2002, \mn@doi [\aap] {10.1051/0004-6361:20011817}, \href {http://adsabs.harvard.edu/abs/2002A%26A...385..337T} {385, 337}

\bibitem[\protect\citeauthoryear{{Tolstoy}, {Hill}  \& {Tosi}}{{Tolstoy} et~al.}{2009}]{Tolstoy2009}
{Tolstoy} E.,  {Hill} V.,   {Tosi} M.,  2009, \mn@doi [\araa] {10.1146/annurev-astro-082708-101650}, \href {https://ui.adsabs.harvard.edu/abs/2009ARA&A..47..371T} {47, 371}

\bibitem[\protect\citeauthoryear{{Tomczak} et~al.,}{{Tomczak} et~al.}{2014}]{Tomczak2014}
{Tomczak} A.~R.,  et~al., 2014, \mn@doi [\apj] {10.1088/0004-637X/783/2/85}, \href {https://ui.adsabs.harvard.edu/abs/2014ApJ...783...85T} {783, 85}

\bibitem[\protect\citeauthoryear{{Trujillo} et~al.,}{{Trujillo} et~al.}{2021}]{Trujillo2021}
{Trujillo} I.,  et~al., 2021, \mn@doi [\aap] {10.1051/0004-6361/202141603}, \href {https://ui.adsabs.harvard.edu/abs/2021A&A...654A..40T} {654, A40}

\bibitem[\protect\citeauthoryear{{Watkins}, {Salo}, {Kaviraj}, {Collins}, {Knapen}, {Venhola}  \& {Rom{\'a}n}}{{Watkins} et~al.}{2023}]{Watkins2023}
{Watkins} A.~E.,  {Salo} H.,  {Kaviraj} S.,  {Collins} C.~A.,  {Knapen} J.~H.,  {Venhola} A.,   {Rom{\'a}n} J.,  2023, \mn@doi [\mnras] {10.1093/mnras/stad654}, \href {https://ui.adsabs.harvard.edu/abs/2023MNRAS.521.2012W} {521, 2012}

\bibitem[\protect\citeauthoryear{{Weaver} et~al.,}{{Weaver} et~al.}{2022}]{Weaver2022}
{Weaver} J.~R.,  et~al., 2022, \mn@doi [\apjs] {10.3847/1538-4365/ac3078}, \href {https://ui.adsabs.harvard.edu/abs/2022ApJS..258...11W} {258, 11}

\bibitem[\protect\citeauthoryear{{Weigel}, {Schawinski}  \& {Bruderer}}{{Weigel} et~al.}{2016}]{Weigel2016}
{Weigel} A.~K.,  {Schawinski} K.,   {Bruderer} C.,  2016, \mn@doi [\mnras] {10.1093/mnras/stw756}, \href {https://ui.adsabs.harvard.edu/abs/2016MNRAS.459.2150W} {459, 2150}

\bibitem[\protect\citeauthoryear{{Zamojski} et~al.,}{{Zamojski} et~al.}{2007}]{Zamojski2007}
{Zamojski} M.~A.,  et~al., 2007, \mn@doi [\apjs] {10.1086/516593}, \href {https://ui.adsabs.harvard.edu/abs/2007ApJS..172..468Z} {172, 468}

\bibitem[\protect\citeauthoryear{{Zarattini} \& {Aguerri}}{{Zarattini} \& {Aguerri}}{2025}]{Zarattini2025}
{Zarattini} S.,  {Aguerri} J. A.~L.,  2025, \mn@doi [arXiv e-prints] {10.48550/arXiv.2504.02026}, \href {https://ui.adsabs.harvard.edu/abs/2025arXiv250402026Z} {p. arXiv:2504.02026}

\bibitem[\protect\citeauthoryear{{de los Reyes} et~al.,}{{de los Reyes} et~al.}{2024}]{Delosreyes2024}
{de los Reyes} M. A.~C.,  et~al., 2024, \mn@doi [arXiv e-prints] {10.48550/arXiv.2409.03959}, \href {https://ui.adsabs.harvard.edu/abs/2024arXiv240903959D} {p. arXiv:2409.03959}

\bibitem[\protect\citeauthoryear{{van den Bergh}}{{van den Bergh}}{1959}]{vandenbergh59}
{van den Bergh} S.,  1959, Publications of the David Dunlap Observatory, \href {https://ui.adsabs.harvard.edu/abs/1959PDDO....2..147V} {2, 147}

\makeatother
\end{thebibliography}


\appendix

\section{Mass function comparison taking into account underestimates of SED-fitted stellar masses}
\label{app:mass_function_offset}

As noted in Section \ref{sec:missing_dwarfs}, the stellar masses derived from SED fitting underestimate the true stellar masses of galaxies, with the average underestimate in the dwarf regime likely to be $\sim$0.2 dex. In Figure \ref{fig:mf_appendix_general}, we show a version of Figure \ref{fig:mf_general} where the original observed mass functions (shown in grey) have been {\color{black} shifted} by 0.2 dex in the dwarf regime (shown in red and blue). While Figure \ref{fig:mf_general} already indicates consistency between data and theory, Figure \ref{fig:mf_appendix_general} illustrates the point that raising the observed mass function by $\sim$0.2 dex in the dwarf regime would result in good agreement between the observations and the ensemble of theoretical simulations explored in this study.

\begin{figure}
\center
\includegraphics[width=\columnwidth]{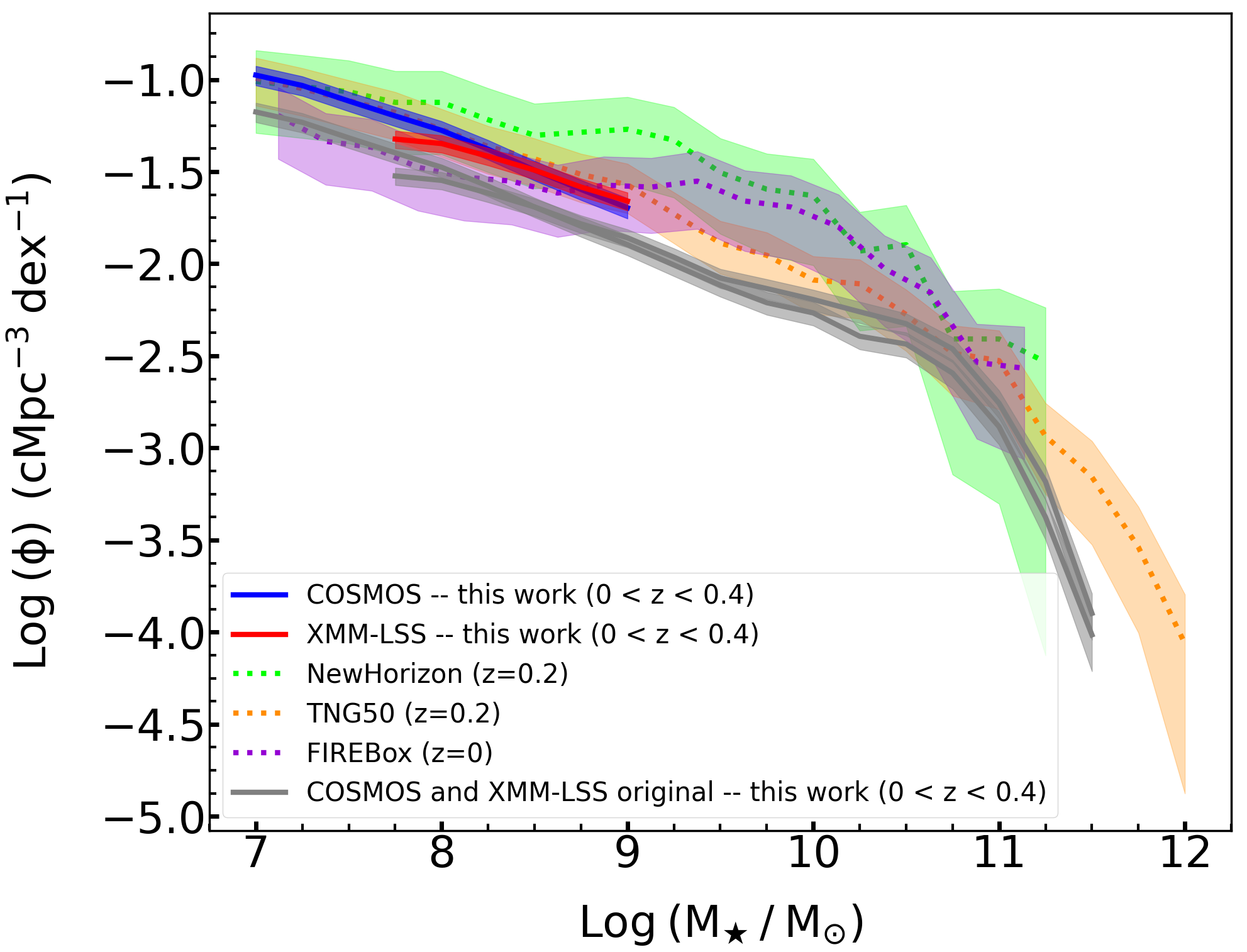}
\caption{The observed galaxy stellar mass function in the nearby Universe in the COSMOS and XMM-LSS fields compared to theoretical mass functions from three high resolution cosmological hydrodynamical simulations (\textsc{NewHorizon}, TNG50 and \textsc{FIREbox}). The original observed mass functions in the COSMOS and XMM-LSS fields, from Figure \ref{fig:mf_general}, are shown in grey. The observed mass functions in red and blue show the original mass functions, in the dwarf regime only, {\color{black} shifted} by 0.2 dex, the approximate offset induced by the fact that SED-fitted stellar masses are typically underestimated in this regime (see text in Section \ref{sec:missing_dwarfs} and Appendix \ref{app:mass_function_offset} for more details).} 
\label{fig:mf_appendix_general}
\end{figure}


\section{Double Schechter fits and associated parameters}

\begin{table*}
	\centering
    \setlength{\tabcolsep}{12pt}
	\caption{Best-fit parameters and their associated uncertainties from double Schechter fits to the mass functions in the COSMOS and XMM-LSS fields in different environments. Galaxies in the lower tercile in the distance from LSS are defined as those that reside in the lowest tercile (0$^{\rm th}$ -- 33$^{\rm rd}$ percentile values) of both the distances to nodes and the distances to filaments. Similarly, galaxies in the middle and upper terciles in the distance from LSS are defined as those that reside in the middle (33$^{\rm rd}$ -- 66$^{\rm th}$ percentile values) and upper (66$^{\rm th}$ -- 100$^{\rm th}$ percentile values) terciles of both the distances to nodes and the distances to filaments respectively. Galaxies in the lower tercile in the distance from LSS are the ones nearest to LSS, while galaxies in the upper tercile in the distance from LSS are furthest away from LSS.}
	\label{tab:schechter_params}
	\begin{tabular}{ccccccc} 
		\hline
		Field & Environment &  Log$_{10}$ M$^\star$ & $\alpha_1$ & $\Phi_1$ $\times$ 10$^{-3}$ & $\alpha_2$ & $\Phi_2$ $\times$ 10$^{-3}$ \\
        & & (M$_\odot$) & & (Mpc$^{-3}$ dex$^{-1}$) & & (Mpc$^{-3}$ dex$^{-1}$) \\
		\hline
		\multirow{3}{5em}{\centering COSMOS} & Lower tercile in dist. from LSS & 10.96 $\pm$ 0.06 & -1.24 $\pm$ 0.03 & 2.37 $\pm$ 0.72 & -0.68 $\pm$ 0.31 & 2.39 $\pm$ 0.83 \\
		& Middle tercile in dist. from LSS & 10.59 $\pm$ 0.07 & -1.34 $\pm$ 0.02 & 1.68 $\pm$ 0.22 & 1.01 $\pm$ 0.35 & 1.26 $\pm$ 0.25 \\
		& Upper tercile in dist. from LSS & 10.53 $\pm$ 0.07 & -1.40 $\pm$ 0.02 & 1.32 $\pm$ 0.18 & 1.41 $\pm$ 0.40 & 0.41 $\pm$ 0.13 \\
		\hline
		\multirow{3}{5em}{\centering XMM-LSS} & Lower tercile in dist. from LSS & 10.95 $\pm$ 0.08 & -1.22 $\pm$ 0.07 & 2.01 $\pm$ 1.01 & -0.56 $\pm$ 0.36 &  3.51 $\pm$ 1.24 \\
		& Middle tercile in dist. from LSS & 10.71 $\pm$ 0.06 & -1.28 $\pm$ 0.03 & 1.93 $\pm$ 0.31 & 0.20 $\pm$ 0.39 & 1.54 $\pm$ 0.34  \\
		& Upper tercile in dist. from LSS & 10.62 $\pm$ 0.08 & -1.34 $\pm$ 0.02 & 1.55 $\pm$ 0.22 & 0.66 $\pm$ 0.37 & 0.91 $\pm$ 0.18  \\
        \hline
	\end{tabular}
\end{table*}

Table \ref{tab:schechter_params} presents the parameters derived from the double Schechter fits to the stellar mass functions discussed in Section \ref{sec:mf_nodes_filaments}. 


\section{Variation of the mass function with distance from nodes and filaments}

\label{app:just_nodes_and_filaments}

Figure \ref{fig:MFdistanceCosmicWeb} shows the observed stellar mass function for different terciles in the distances to filaments (top panel) and nodes (bottom panel) separately. In other words, these panels show versions of Figure \ref{fig:MFcosmicWeb} where the distances to nodes and filaments are considered separately. Figure \ref{fig:MFdistanceCosmicWeb} demonstrates that the differences between galaxies in the different distance terciles are much more significant when considering the distances to filaments than the distances to nodes. Thus, the overall trends seen in Figure \ref{fig:MFcosmicWeb} are driven predominantly by the distances of galaxies from filaments rather than their distances to nodes. 

\begin{figure}
\center
\includegraphics[width=\columnwidth]{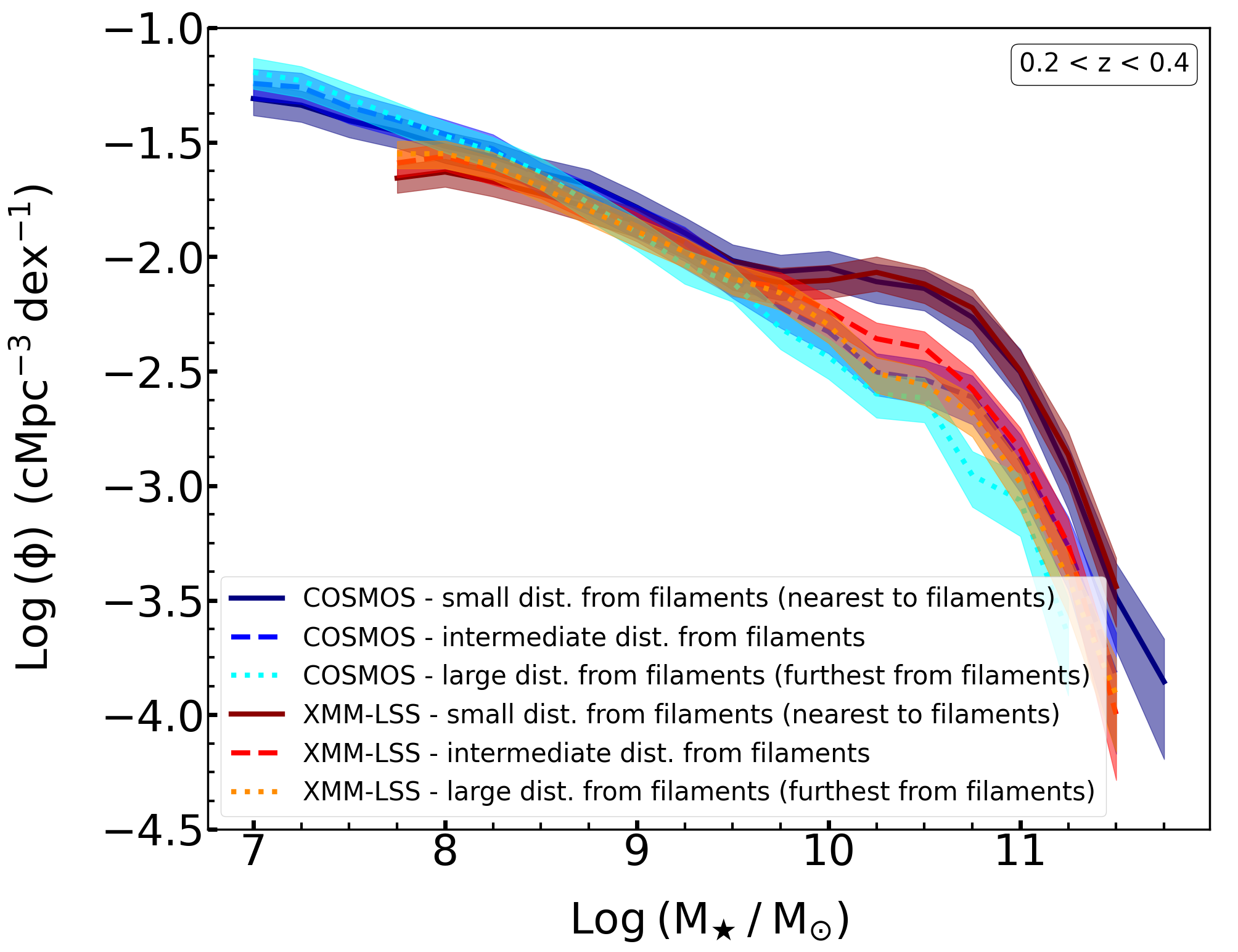}
\includegraphics[width=\columnwidth]{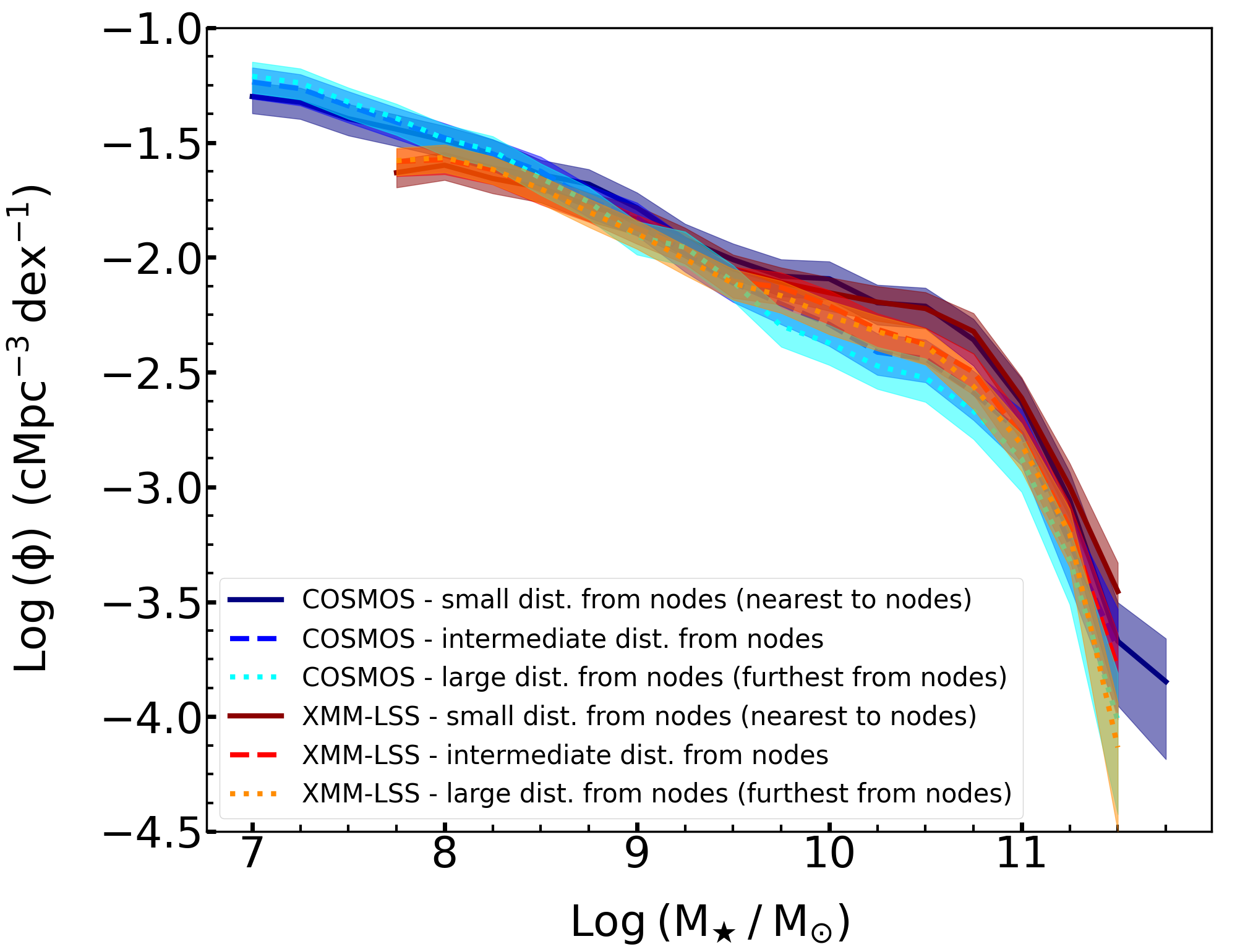}
\caption{The galaxy stellar mass function for different terciles in the distances to filaments (top) and nodes (bottom) separately. These panels show versions of Figure \ref{fig:MFcosmicWeb} where the distances to nodes and filaments are considered separately.}
\label{fig:MFdistanceCosmicWeb}
\end{figure}


\section{SED-fitted stellar masses with and without medium-band filters}
\label{app:medium_band_filters}

In this section we further explain our reasoning for exploring the COSMOS dataset down to stellar masses of 10$^7$ M$_\odot$ but restricting the XMM-LSS dataset to stellar masses above 10$^8$ M$_\odot$. The mass function in COSMOS, derived using both broadband and medium band filters, is shown using the blue curve. The mass function in XMM-LSS, derived using the nine available broadband filters spanning the $u$-band to rest-frame near-infrared ($ugrizyJHK$), is shown using the green curve. The XMM-LSS mass function shows good consistency with the COSMOS mass function down to $M_\star$ $\sim$ 10$^8$ M$_\odot$, after which it shows a downturn which is not shared by its COSMOS counterpart. 

The dashed cyan curve shows the mass function in COSMOS produced by an identical SED fitting procedure to that used to create the blue (XMM-LSS) curve, i.e. using $ugrizyJHK$ only with the medium band filters omitted. Interestingly, this version of the COSMOS mass function also shows a downturn which closely follows that in XMM-LSS (which, recall, does not have any medium band filters). Thus, we speculate that the inconsistency, at $M_\star$ < 10$^8$ M$_\odot$, between the blue curve (which employs SED fitting on broad and medium band filters) and its green and cyan counterparts (which employ SED fitting on broadband filters only) could be driven by the unavailability of medium band filters. This leads to a poorer sampling of the SED that seems to affect very low mass galaxies ($M_\star$ < 10$^8$ M$_\odot$). Due to this, as a precaution, we consider the XMM-LSS dataset only down to $M_\star$ $=$ 10$^8$ M$_\odot$ where it shows good consistency with the COSMOS mass function that is derived using both broad and medium band filters. 

\begin{figure}
\center
\includegraphics[width=\columnwidth]{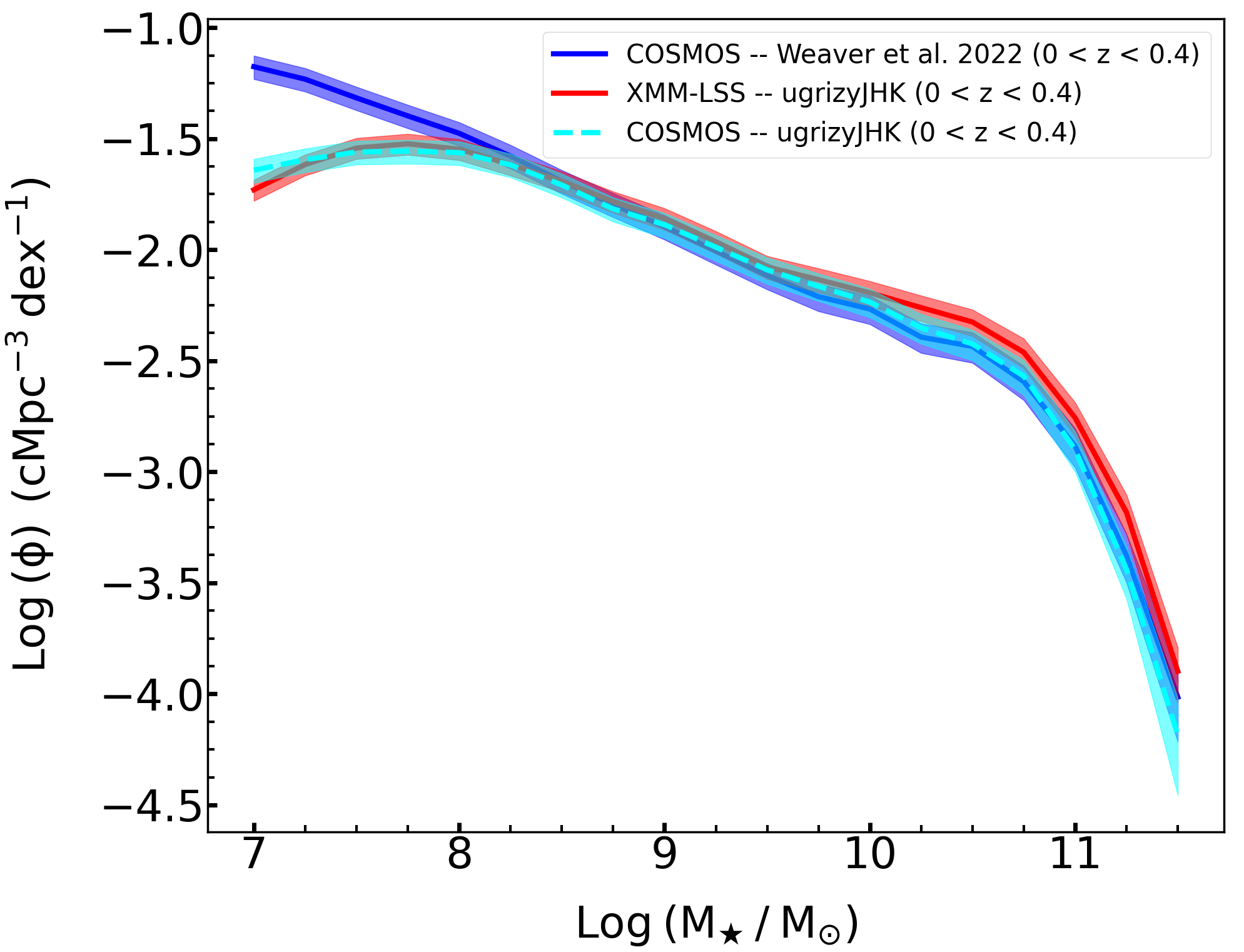}
\caption{The observed galaxy stellar mass functions in the COSMOS and XMM-LSS fields for $z<0.4$ derived using different filter combinations. The blue curve uses stellar masses in COSMOS that are derived using broad and medium band photometry from the COSMOS2020 catalogue. The {\color{black} red} curve uses stellar masses in XMM-LSS that are derived using only broadband $ugrizyJHK$ filters. The dashed cyan curve uses stellar masses in COSMOS that are derived using only broadband $ugrizyJHK$ filters like in XMM-LSS.}
\label{fig:COSMOS_XMM_comp}
\end{figure}


\section{Eddington bias effects on the galaxy stellar mass function}
\label{app:eddington_bias}

In this section we show that the systematic Eddington bias has a negligible influence on the shape of the galaxy stellar mass functions derived in this work. In particular, the COSMOS (blue curve) and XMM-LSS (red curve) stellar mass functions are very similar to the mean of the galaxy stellar mass functions derived using the perturbed stellar masses as described in Section \ref{sec:MonteCarlo}. As shown in Figure \ref{fig:edd_bias}, they are typically within $\sim$0.01 dex until $M_\star$ $\sim$ 10$^{11}$ M$_\odot$, with a maximum deviation of $\sim$0.08 dex at the very massive end of the XMM-LSS mass function ($M_\star$ $\sim$ 10$^{11.5}$ M$_\odot$).

\begin{figure}
\center
\includegraphics[width=\columnwidth]{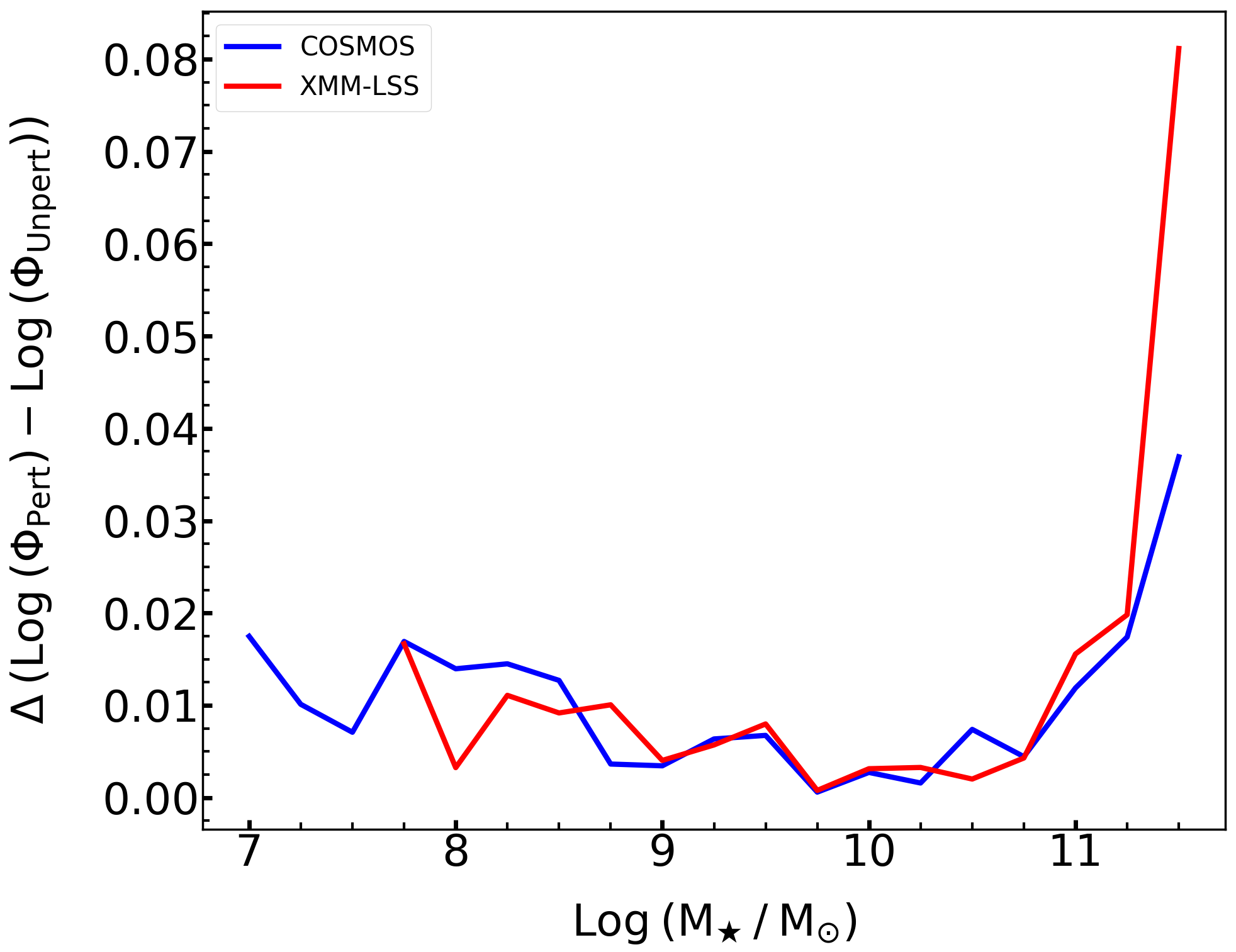}
\caption{ The blue and red curves show the difference between the unperturbed galaxy stellar mass functions in the COSMOS and XMM-LSS fields, respectively, for $z<0.4$ (same as in Figure \ref{fig:mf_general}) and the mean of the galaxy stellar mass function derived using the perturbed stellar masses via the Monte Carlo approach described in section \ref{sec:MonteCarlo}. The perturbed and unperturbed mass functions are denoted using $\Phi_{\rm Pert}$ and $\Phi_{\rm Unpert}$ respectively.}
\label{fig:edd_bias}
\end{figure}


\bsp 
\label{lastpage}
\end{document}